\documentclass[12pt]{article}
\usepackage{epsfig}

\begin{document}

\title{Analysis of zero-frequency solutions of the pion dispersion
equation in nuclear matter}

\author{V. A. Sadovnikova\\
Petersburg Nuclear Physics Institute, \\
Gatchina, St.~Petersburg 188300,
Russia}
\date{}
\maketitle
\begin{abstract}
In this paper we consider instability of nuclear matter which
takes place when the frequencies of the collective excitations
turn to zero.  We investigate collective excitations with pion
quantum numbers $J^\pi=0^-$.  We study the dependence of
zero-frequency solutions of the pion dispersion equation on the
value of the spin-isospin quasiparticle interaction $G'$. The
solutions of the pion dispersion equation describe the different
types of the excitations in the matter, $\omega_i(k)$.  At the
critical density  $\rho=\rho_c$ one of solutions of the definite
type turns to zero: $\omega_{i0}(k_c)$=0. When $\rho>\rho_c$, the
excitations $\omega_{i0}(k)$ become amplified.  It is shown that
there is such a ``transitional" value of $G'=G'_{tr}$ that for
$G'<G'_{tr}$ the zero-frequency solutions belong to the type
$\omega_{sd}$ while for $G'>G'_{tr}$ they pertain to the type
$\omega_c$. The solutions of the type $\omega_{sd}$ correspond
to instability to small density fluctuations of the nuclear
matter at $G'\le -1$.  On the other hand,    $\omega_c$ is
responsible for the ``pion condensation" at $G'\approx 2$.  For
the  stable nuclear matter the branches of solutions
$\omega_{sd}(k)$ and $\omega_c(k)$ are located on the unphysical
sheets of the complex plane of frequency.
\end{abstract}

\section{Introduction}

 In this paper we investigate the excitations of nuclear matter
with the pion quantum numbers.  Our analysis is based on studies
the in-medium pion dispersion equation. Interactions of pions
with baryons in nuclear matter is included in framework of Migdal
model \cite{Mi}.
Solutions of the pion dispersion equation in this model were
considered on the physical sheet of the complex plane of pion
frequency $\omega$.
We expand this analysis to the unphysical sheets of
the Riemann surfaces. We have included first the
unphysical sheets to the analysis in our earlier papers
\cite{DRS,SR}. In \cite{DRS,SR} we studied the branches of
solutions responsible for the pion condensation and the long
wavelength instability and it was shown that they are the
separate branches supplemented to the well-known zero-sound,
pion, and isobar ones.

In this paper we continue to study the solutions of zero-sound
and pion dispersion equations responsible for the instability of
nuclear matter using the retarded pion propagator.  We analyze
zero-frequency solutions of the pion dispersion equation
depending on the value of $G'$ and define to what type
excitations  these zero-frequency solutions refer.  This could
permit us to do conclusion about the character of the phase
transition related to the considered instability.

Except the purely theoretical interest the problem has various
applications.  Investigation of the pion excitations in  nuclear
matter, started long ago, continue to play  an important role in
nuclear physics. Special interest in study of the pionic
many-body degrees of freedom was stimulated by prediction and
investigation of the pion condensation \cite{Mi,SS,BW,OTW}.  In
recent years the study of the pionic many-body degrees of
freedom is related to the investigations of the excited states
 and of phase transitions in nuclear matter in heavy ion
collisions.  Detailed knowledge of in-medium pion dynamics is
essentially important for description of mesons (\cite{Mi,OTW}),
nucleons (\cite{DRS}) and $\Delta$-isobars \cite{MSG},
\cite{HR}.  Analysis of the dilepton
production from $\pi^+-\pi^-$ annihilation requires the
knowledge of the pion dispersion in hot and dense matter \cite
{HR,GK,XX}.

To study of the pion dynamics in heavy ion collisions we need
the relativistic description of the pion self-energy at the
finite temperatures. Solutions of the relativistic pion
dispersion equations were presented in papers \cite{LN,AM}. A
pion self-energy with the correct nonrelativistic limit was
obtained in papers \cite{LP,Lutz}.

When considering  the pion dynamics in nuclear matter at high
densities and temperatures, it is important to have as a "point
of reference" a reliable description of the nuclear matter
excitations with pion quantum numbers in nonrelativistic limit.
Such description have been obtained in the pioneering papers of
Migdal and his group \cite {Mi}, followed by the numerous papers
\cite {BW,OTW,BBN,GOS,CE,DG}.

In this paper we study the solutions of the nonrelativistic pion
dispersion equations in symmetric nuclear matter at zero
temperature following to \cite{Mi}.
The aim of this  paper is to study the solutions
with zero frequency $\omega$=0  depending
on the value of the coupling constant  $G'$. These solutions
characterize the  stability boundary (in this model it is the
boundary on the density).  In this paper it is shown that at
different values of $G'$ the zero-frequency solutions belong to
the different types of excitations.

The value of $G'$ is considered in the interval $-1\leq G'\leq
2$.  When we change  $G'$, the branches of solutions are changed
as well:  certain solutions go over from the physical to
 unphysical sheets and vice verse. To identify the solutions on
the unphysical sheets it is important to know the solutions on
the physical ones.  The branches of solutions which are obtained
in the present paper reproduce the results of \cite{Mi,CC,PR} on
the physical sheet of the complex plane of $\omega$.  To do
comparison with other papers the simple model with the constant
effective quasiparticle interaction is very useful.  However,
the pion condensation in this model emerges at too low density
and this is not consistent with the results of investigations,
\cite{Mi}.

 It is well known that the solutions of the pion dispersion
equations, $\omega_i(k)$, describe the different types of
excitations in the nuclear matter. They are:  spin-isospin
zero-sound wave, $\omega_s(k)$, pion wave, $\omega_\pi(k)$,
isobar wave, $\omega_\Delta(k)$, and others \cite{Mi,BW,OTW,EW}.
The appearance of the solution with $\omega$=0 means that at the
certain values of nuclear density $\rho=\rho_c$ and wave vector
$k=k_c$ the frequency of a definite type of excitations $i_0$
 turns to zero:  $\omega_{i_0}(k_c)$=0.  If $\rho >\rho_c$ one
 obtains $Im\omega_{i_0}(k)>0$, and the amplified solution takes
place.  This signals the instability of nuclear matter.  The
change of $G'$ causes the changes of the
values of $\rho_c$ and $k_c$.  Moreover, at a special situation
zero-frequency solution passes to another type of excitations
$i_1$:  $\omega_{i_1}(k_c)$=0. It was shown in \cite{LaLi} that
the phase transition is determined by the type of those
excitations which become unstable. Thus it is important to know
to what  type of excitations the solutions with $\omega$=0
belongs.

In Fig. 1 we show the values of critical densities
$\rho_{c}$ and wave vectors
$k_{c}$ for which the pion dispersion equation
\begin{equation}\label{1}
 D^{-1} = \omega^2-k^2-m^2_\pi-\Pi^R(\omega,k)\ =\ 0\ ,
\end{equation}
has solutions with $\omega$=0 at any value of $G'$.  We are
interested in solutions which satisfy  additional restriction:
$\left(\frac d{dk}\omega(k)\right)_{k=k_c}$=0.
In Eq.~(1) $\Pi^R(\omega,k)$
is the pion self-energy part (retarded polarization operator).
The pion self-energy is formed on the basic of particle-hole and
isobar-hole loops renormalized due to the effective interactions of
quasiparticles: $G'_{NN},$ $G'_{N\Delta}$ and $G'_{\Delta\Delta}$.
(The effective constant $G'$ is regarded as
 $G'_{NN}$ through the paper.) In Fig.~1 the critical density $\rho_c$
and wave vector $k_c$ are presented for three models of the pion
self-energy. The matter is stable at $\rho<\rho_c$ for the every
model.
The physical variables which are related
to the pion propagator $D$ diverge along the curves in Fig.~1.
This corresponds to the phase transition \cite{LaLi} in nuclear
matter. At $\rho<\rho_c$, the nuclear matter is considered as a
normal Fermi liquid, but at $\rho>\rho_c$ there is   phase
transitions into  new state which may include the standing
spin-isospin wave or "condensates of excitation".

Let us introduce the useful definitions. The waves
$\omega_s(k)$, $\omega_\pi(k)$, $\omega_\Delta(k)$, and others
we name branches of solutions. The set of branches of solutions
of the same type for the different values of $G'$ composes a
family ($\omega_s$, $\omega_\pi$, $\omega_\Delta$, and others).
When we say about zero-frequency solutions or $\omega$=0
solutions, we mean that the condition
$\left(\frac d{dk}\omega(k)\right)_{k=k_c}$=0 is satisfied in
this point, as well.

Return to Fig.~1.
There are known two reasons for instability in the framework of
Migdal theory \cite{Mi} in the nonrelativistic symmetric cold
nuclear matter at not too high densities. In paper
\cite{Pom} it was shown that Fermi liquid becomes unstable when
the attractions between the quasiparticles is too strong.  In
spin-isospin channel this corresponds to $G'<-1$ \cite{Mi}.  In
Fig.~1 we see that there are not solutions with $\omega$=0 at
 $G'<-1$.  This is because of the matter is unstable at the any
density.  Another
region of $G'\approx 2$ corresponds to the instability in
respect to  transition into a new ground state containing
condensate of pions \cite{Mi}.
In the present paper the main attention is paid to study that
families of solutions which are responsible for instability
along the curves in Fig.~1, from long wavelength instability to pion
condensation.

We start with the analysis of the zero-sound dispersion equation for
spin-isospin excitations with quantum numbers  $J^\pi=0^-$. This
equation is closely related to the pion dispersion equation. It was
shown in paper \cite{LaLi,PN} that for the case of strong
attraction between the quasiparticles ($G'<-1$) the zero-sound
dispersion equation has a growing with time solution which
corresponds to  instability in response of small-amplitude, long
wavelength density fluctuations. For  $G'>-1$ the unstable solutions do
not appear \cite{CC,PR}.

In this paper the family of solutions responsible for the long
wavelength instability is built. We denote this family as
$\omega_{sd}$. The index ``sd" means ``spinodal decomposition".
It is known \cite{LaLi}, that the thermodynamic stability
condition requires  the partial derivative of the pressure with
respect to the volume to be negative, i.e.,
$\left(\frac{\partial P}{\partial V}\right)_T < 0$. The process
which takes place in matter, when this condition is broken, is
known as a spinodal decomposition \cite{MS}. It is shown in
\cite{LaLi}, that there is a relation between the effective
quasiparticle interaction and the partial derivative
$$
\frac{\partial P}{\partial V} = -\frac{N}{V^2} \frac{p_F^2}{3m}
(1 + F_0),
$$
$N$ is the number of particles, $F_0$ is the effective scalar-isoscalar
quasiparticle interaction. The stability condition is broken at $F_0 <
-1$.  For the zero-sound dispersion equation this corresponds to
the appearance of the imaginary amplified solutions. It looks reasonable to
use the notation $\omega_{sd}(k)$ for such solutions. Here we use the
same notation for the analogous solutions, which appear in the
spin-isospin channel at $G'<-1$. Moreover, as it is shown below,
 at $G' < 0$ all the solutions belong to the same family. We use
the notation $\omega_{sd}(k)$ for them as well.
The branches $\omega_{sd}(k)$
are imaginary and can be found for all values $G' < 0$.

Now we turn to the pion dispersion equation. In the channel of
excitations with the pion quantum numbers the
particle-hole interaction can be modified by taking into account
the pion-nucleon interactions \cite{Mi}. As a result, collective
and particle-hole  excitations are renormalized due to
interaction with pions and, vice versa, the pion excitations are
changed due to interaction with particle-hole ones in the
nuclear matter. It was predicted \cite{Mi,SS}, that the account
of the pion-nucleon interaction is responsible for the phase
transition into the ground state with the pion condensation.  In
Fig.~1 it is demonstrated that due to the pion-nucleon
interaction we obtain the phase transition not only for $G' <
-1$ but for all (considered) values of $G'$.

For the pion dispersion equation (\ref{1}) we can find the branch
of solutions corresponding to  $\omega_{sd}(k)$. We denote it as
$\omega^\pi_{sd}(k)$. The correspondence is established in the
following way. It will be seen from the form of the pion self-energy
that $\omega_{sd}(k)$ are the poles of $\Pi(\omega,k)$, and for
each value of  $k$  the frequency $\omega^\pi_{sd}(k)$
tends to  $\omega_{sd}(k)$ as
pion-nucleon coupling constant $f_{\pi NN}\rightarrow 0$.

We  show also that  there is  a
special value of $G'$, denoted by
$G_{tr}$ ($|G'_{tr}|\ll 1$). When $G' < G'_{tr}$, the solutions with
$\omega$=0 are on the branches $\omega^\pi_{sd}(k)$.
When $G'>G'_{tr}$, the type of the family  carrying
zero-frequency solutions is changed. We shall refer to $G'_{tr}$ as
"transition value".

When $G'>G'_{tr}$, zero-frequency solutions pass to the branches
of solutions of another family.  It is denoted as $\omega_c$
(``condensate").  The solutions of (1) referred to this family,
$\omega_c(k)$, are identified in the following way. In papers
\cite{Mi} it was shown that among the solutions of the pion
dispersion equation the imaginary solutions emerge under
special conditions.  Instability of matter, related to these
imaginary solutions  is interpreted as a phase transitions into
a new ground state containing the pion condensate. The branch
$\omega_c(k)$ reproduces the imaginary solutions presented in
\cite{Mi} on the physical sheet of $\omega$.  It is
demonstrated that $\omega_c$ is an independent family of
solutions along with $\omega_\pi$, $\omega_s$, $\omega_\Delta$
or $\omega^\pi_{sd}$ and it is responsible for pion
condensation.

In Sect.~2 the expressions for the retarded pion self-energy are
presented and the singularities on the complex
plane are investigated. We consider the structure of the
Riemann surfaces that are defined by the form of the pion self-energy.
The unphysical sheets are described.

In Sect.~3 the family $\omega_{sd}$ is considered for
the zero-sound and pion dispersion equations. It is shown how
zero-frequency solutions   $\omega_{sd}(k_c)$=0 at
$\rho=\rho_c$ (Fig.~1) come these branches.

In Sect.~4 the different types of solutions of the pion dispersion
equation  are considered on the physical and unphysical sheets.
It is shown that there is an independent family of solutions
$\omega_c$ that is related to the pion condensation.

In Sect.~5 we determine the value of   $G'_{tr}$  and show how
the  solution with $\omega$=0 pass from the family
$\omega^\pi_{sd}(k)$  to $\omega_c(k)$ when $G'= G'_{tr}$.

In this paper we solve the dispersion equations  in the symmetric
nuclear matter with zero temperature. The equilibrium density is
$\rho_0$, $p_{F0}$=0.268~GeV. During the computations, the
effective mass of quasiparticles is assumed to be  $m$=0.8$m_0$,
while $m_0$=0.94~GeV is the vacuum nucleon mass. The pion mass
is $m_\pi$=0.14~GeV, the isobar mass
$\Delta=1.23$~GeV (the isobar width is neglected).
Following to the paper \cite{Mi}, the effective nucleon quasiparticle
interaction can be written as
$$
{\cal F}(\vec k_1,\vec \sigma_1,\vec
\tau_1;\vec k_2,\vec \sigma_2,\vec \tau_2)=C_0( F(\vec k_1,\vec k_2)
+F'(\vec k_1,\vec k_2)(\vec \tau_1\vec \tau_2) +G(\vec k_1,\vec
k_2)(\vec \sigma_1\vec \sigma_2)
$$
$$
+ G'(\vec k_1,\vec k_2)(\vec
\sigma_1\vec \sigma_2) (\vec \tau_1\vec\tau_2)),
$$
where $\vec \tau,
\vec \sigma$ are the Pauli matrices in the isospin and spin spaces. The
factor $C_0=N_0^{-1}$, where $N_0$ is a density of states for  two
sorts of nucleon on the Fermi surfaces, $N_0=2p_Fm/\pi^2$;
$F, F', G, G'$ are the Landau-Migdal dimensionless parameters.
 In this paper we use
a simple model where parameters are taken to be a constant:
$G'(\vec k_1,\vec k_2)=G'(p_F,p_F)\equiv G'$.
The Landau-Migdal parameter $G'$ found from the analysis of the
nuclear experimental data is  $G'\sim 1.5-1.7$ \cite{BBN,
SPWW}.  This parameter describes the short-range correlations between
the nucleons with a range $\sim \frac1{m_0}$:
$V_{corr}=\frac{f^2}{m_\pi^2}g'(\vec \sigma_1\vec \sigma_2)(\vec
 \tau_1\vec\tau_2)$. The relation $G'=\frac1{C_0}\frac{f^2}{m_\pi^2}g'$
sets the correspondence of the Landau-Migdal parameters to the effective
correlative parameters $g'$ \cite{LN,AM,LP}.

The Landau-Migdal
parameter $G'=G'_{NN}$ corresponds to rescattering of
$NN\leftrightarrow NN$ quasiparticles.  The parameters
$G'_{N\Delta}$ and $G'_{\Delta\Delta}$ determine the
 rescattering of $N\Delta\leftrightarrow NN$ and
$N\Delta\leftrightarrow N\Delta$.
The parameters $G'_{N\Delta}$ and $G'_{\Delta\Delta}$
 are not changed in the course of
paper  and are taken equal to $G'_{N\Delta}$=0.4 and
$G'_{\Delta\Delta}$=1.6 \cite{Mi}.

In the paper we take the pion-nucleon coupling constant $f_{\pi
NN}=1$ and the pion-nucleon-isobar coupling constant $f_{\pi
N\Delta}=2$.

\section{The pion self-energy $\Pi(\omega,k)$}
In this Section we present the expressions for the pion self-energy
in nuclear matter. These expressions are well known
\cite{Mi,EW,PN,FW,DM}. We do an analytical continuation in  $\omega$
of the polarization operator $\Pi(\omega,k)$ to
 the unphysical sheets of the complex plane.
The search of zero-frequency solutions is  related to
investigation of the solutions both on the physical and
unphysical sheets.

We consider excitations in nuclear matter which are described by the
solutions of the pion dispersion equation (1).
The causality requirement leads to the known interpretation of the
solutions of the dispersion equation in the complex plane of $\omega$.
The stable excitations are described by the solutions with real
frequencies, placed on the real axis. The damping solutions (resonances)
are placed on the unphysical sheets under the cuts, which the
polarization operator has on the real axis.
This is Landau damping which is related to overlapping of the frequency
of excitations with the frequency of free particle-hole pairs (which
give the imaginary part in the pion self-energy).  We can introduce the
damping of solutions in another way, with the help of the isobar width
(taken as the experimental constant, $\Gamma$=115~MeV) or quasiparticle
damping constant, for example. In this case the real excitations are
displaced to the lower half plane of the physical sheet \cite{DRS,SR}.
We do not consider this damping henceforth.

At certain conditions (large density, a
strong attraction between particles) the imaginary solutions
arise in the upper half-plane of the physical sheet.  They
correspond to the excitations with amplify amplitude. Such
excitations correspond to an unstable regime. We find the
corresponding solutions with negative imaginary part in the
lower half-plane of the physical sheet, as well.

 The amplified solutions in the spin-isospin channel
(as well as all the solutions with $Re\omega_i>0$)
are interpreted, following \cite{Mi,BaW}, as the
frequencies of $\pi^+$-type excitations.  While the solutions in the
lower half-plane
(as well as all the solutions with $Re\omega_i<0$)
are considered as $\pi^-$-type excitations. Their frequencies have to
be taken with an opposite sign in respect to the solutions.
 For the physical excitations of
$\pi^0$-type we put into correspondence the solutions with
$Re\omega_i>0$.  In symmetrical nuclear matter with $N=Z$ all three
types of solutions, i.e., $\pi^+, \pi^-, \pi^0$ are identical.
In the asymmetric nuclear matter the
solutions carrying  different charges are transformed in a different
way with the change of the parameter of asymmetry and of the density.  The
method used in this paper permits to follow these transformations
 from  symmetric up to  neutron nuclear matter.

Pion self-energy in nuclear matter is  due to $S$- and
$P$-interaction of the pions and nucleons. The total pion self-energy
is thus a sum of a scalar and vector terms:
\begin{equation}\label{1f}
\Pi(\omega,k)\ =\ \Pi_S(\omega,k)+\Pi_P(\omega,k)\ .
\end{equation}
The main results of this paper are obtained for
$\Pi(\omega,k)=\Pi_P(\omega,k)$
operator. One can consider some models for $\Pi_S(\omega,k)$ and they
give the expressions like $\Pi_S = const\cdot\rho$ without $\omega$ and
$k$ dependence \cite{DRS,EW}. The influence  of $\Pi_S$ on
the final results has the quantitative character and is discussed in
Appendix A.

\subsection{$P$-wave part of the pion self-energy, $\Pi_P(\omega,k)$}

Following the papers \cite{Mi,PN,FW} we write $\Pi_P(\omega,k)$
as a sum of the
nucleon and isobar parts
\begin{equation}\label{2f}
\Pi_P(\omega,k)\ =\ \Pi_N(\omega,k)+\Pi_\Delta(\omega,k)\ .
\end{equation}
The operators $\Pi_N(\omega,k)$
and $\Pi_\Delta(\omega,k)$ can be expressed through
the lowest order polarization insertions
$\Pi^0_N(\omega,k)$ and $\Pi^0_\Delta(\omega,k)$
which can be presented as the integrals over the particle-hole ($\Phi_N$) and
isobar-hole loops ($\Phi_\Delta$). The vertices of these loops
are described by the pion-nucleon and isobar-nucleon-pion coupling constants
$f_{\pi NN}$, $f_{\pi N\Delta}$ and by the form
 factors  which take the nonzero baryon size into account,
$d_B(k^2)=\frac{\Lambda_B^2-m_\pi^2}{\Lambda_B^2+k^2}$,
where $B=N, \Lambda$ and $\Lambda_N$=0.667~GeV,
$\Lambda_\Delta$=1.0~GeV.  Polarization operators
$\Pi^0_N(\omega,k)$,
$\Pi^0_\Delta(\omega,k)$ can be written as:
\begin{equation}\label{3f}
\Pi^0_N(\omega,k) =
-4(\frac{f_{\pi NN}}{m_\pi})^2 k^2 d^2_N(k^2)\Phi_N(\omega,k),
\end{equation}
\begin{equation}\label{4f}
\Pi^0_\Delta(\omega,k)
= -\frac{16}9(\frac{f_{\pi N\Delta}}{m_\pi})^2 k^2
d^2_\Delta(k^2)\Phi_\Delta(\omega,k).
\end{equation}
Then for $\Pi_N$, $\Pi_\Delta$ we have the expressions \cite{DM}:
$$
\Pi_N(\omega,k) = \Pi^0_N(\omega,k)
\frac{1+C_0(\alpha(k)G'_{N\Delta} -G'_{\Delta\Delta})
\frac{-16}9\Phi_\Delta(\omega,k)}{E(\omega,k)} ,
$$
\begin{equation}\label{5f}
\Pi_\Delta(\omega,k)
= \Pi^0_\Delta(\omega,k)\frac{1+C_0(\frac1{\alpha(k)}G'_{N\Delta} -
G'_{NN})(-4)\Phi_N(\omega,k)}{E(\omega,k)}.
\end{equation}
In this equation $\alpha(k)=(\frac43f_{\pi N\Delta}
d_\Delta(k^2))/(2f_{\pi NN}d_N(k^2))$. In Eqs.~(\ref{5f}) we see
that $\Pi_N(\omega,k)$
is $\Pi^0_N(\omega,k)$ with one vertex renormalized due to
nucleon-hole and isobar-hole interaction.  The denominator $E(\omega,k)$
has a form
\begin{eqnarray}
E(\omega,k)&=&1+C_0\left(4G'_{NN}\Phi_N(\omega,k)
+ \frac {16}9G'_{\Delta\Delta}\Phi_\Delta(\omega,k)\right.
\nonumber\\
&&\left. \quad +4\frac{16}9C_0(G'_{NN}G'_{\Delta\Delta} -
 G'^2_{N\Delta})\Phi_N(\omega,k)\Phi_\Delta(\omega,k)\right)
\label{6f}
\end{eqnarray}
and the condition
\begin{equation}\label{7}
E(\omega,k)=0
\end{equation}
gives a zero-sound dispersion equation. Solutions of this equation
describe the excitations in nuclear matter, if we do not take into
account the renormalization  due to the pion-baryon interaction.
Equation (1) has the solutions corresponding to the solutions of
Eq.~(\ref{7}). Thus at $G'_{NN}<0$, Eq.~(\ref{7}) has the
solutions $\omega_{sd}(k)$ (which will be shown and discussed
below), while at $G'_{NN}>0$  there are zero-sound solutions
$\omega_s(k)$ \cite{CC,PR,LaLi}. These solutions are the poles
of $\Pi_N(\omega,k)$ and $\Pi_\Delta(\omega,k)$ (\ref{5f}). The
equation (1) has the corresponding solutions:
$\omega^\pi_{sd}(k)$ and $\omega^\pi_s(k)$.  It can be seen from
the structure of Eq.~(1) that when  $f_{\pi NN}$ and $f_{\pi
N\Delta}$ approach zero points, the branches
$\omega^\pi_{sd}(k)$ ($\omega^\pi_s(k)$) approach to the poles
of polarization operator $\Pi(\omega,k)$, i.e., $\omega_s(k)$
($\omega_{sd}(k)$).

\subsection{Extension of $\Pi^0_N(\omega,k)$ in $\omega$
to the unphysical sheets of the complex plane}

The causal polarization operator $\Pi^0_N(\omega,k)$
can be presented as a sum of two terms,
 the first  term corresponds to  excitation and the second to the
annihilation of the particle-hole pairs.
We want to study the excitations in nuclear matter
 which appear as a response on an external perturbation and we  use
the retarded operators in Eq.~(1). For the real $\omega$ they are
determined as \cite{LaLi,FW}
$$
\Pi^{0R}(\omega,k) = (Re + i\
sign(\omega) Im) \Pi^0(\omega,k)
$$
and $\Phi^R_N(\omega,k)$ is expressed through the integrals
\cite{FW}
\begin{equation}\label{9a}
\Phi^R_N(\omega,k) =
\frac{-1}{(2\pi)^3}\int d^3p\left[ \frac{\theta(|\vec p+\vec
k|-p_F)\theta(p_F-p)} {\omega+\frac{p^2}{2m}-\frac{(\vec p+\vec
k)^2}{2m}+i\eta} - \frac{\theta(p-p_F) \theta(p_F-|\vec p+\vec
k|)} {\omega+\frac{p^2}{2m}-\frac{(\vec p+\vec k)^2}{2m}+i\eta}
\right] .
\end{equation}
The function $\Phi^R_N(\omega,k)$ can be written as
\begin{equation}\label{9}
\Phi_N(\omega,k) =
\phi_N(\omega,k) + \phi_N(-\omega,k).
\end{equation}
The first integral in (\ref{9a}) corresponds to $\phi(\omega,k)$ and
the second one to $\phi(-\omega,k)$.
Equation (\ref{9a}) can be used to obtain  a condition
   which relates the values of $\Phi^R_N(\omega,k)$
in the two points of the complex plane \cite{LaLi}
\begin{equation}\label{1g}
\Phi^R_N(-\omega^*,k) = \Phi^{R*}_N(\omega,k).
\end{equation}
The same equation can be written for $\Phi^R_\Delta$ and, consequently,
for operators in Eqs.~(\ref{2f})-(\ref{6f}).  Henceforth only the
retarded operators are used and  index $R$ is omitted.

We write the expression for $\Phi_N(\omega,k)$  in such a way that it
does not contain the overlapping logarithm cuts (compare
with \cite{PN,FW}):
\begin{equation}\label{10}
\phi_N(\omega,k)=
\frac{m}{k} \frac1{4\pi^2}
\left(
\frac{-\omega m+kp_F}2-\omega m \ln\left(\frac{\omega m}{\omega
m-kp_F+\frac12k^2}\right)\right.
\end{equation}
$$
+\left.
\frac{(kp_F)^2-(\omega m-\frac12k^2)^2}{2k^2} \ln\left( \frac{\omega
m-kp_F-\frac12k^2}{\omega m-kp_F+\frac12k^2}\right)\right)
$$
for $0\le k\le2p_F$. Each of logarithms has its own region of
integration over $p$ and over the angle between $\vec p$ and $\vec k$
in (\ref{9a}).  These regions are shown in \cite{FW}.

At  $k\ge2p_F$,  $\phi_N(\omega,k)$ is the Migdal's function
\cite{Mi,EW}
\begin{equation} \label{11}
\phi_N(\omega,k) =\ \frac1{4\pi^2}\
\frac{m^3}{k^3}
\left[\frac{a^2-b^2}2 \ln\left(\frac{a+b}{a-b}\right)-ab\right]\,
\end{equation}
where $a=\omega-(k^2/2m)$, $b=kp_F/m$.

Function $\Phi_\Delta(\omega,k) = \phi_\Delta(\omega,k) +
\phi_\Delta(-\omega,k)$, where $\phi_\Delta(\omega,k)$ is determined by
 (\ref{11}) with $a=\omega-(k^2/2m) + (m_\Delta-m)$.

Let us consider the cuts of $\Pi^0_N(\omega,k)$ in the complex plane
of $\omega$. It is clear that, for $k\le2p_F$, there are two cuts (we
denote them by $I$ and $II$). The cuts originated by the
logarithmic terms of (\ref{10}) are at
\begin{equation} \label{12}
I: 0\ \le\ \omega\ \le\
\frac{kp_F}{m} -\frac{k^2}{2m}\,
\end{equation}
$$
 II:
\frac{kp_F}{m}-\frac{k^2}{2m}\ \le\ \omega\ \le\
\frac{kp_F}{m}+\frac{k^2}{2m}\ .
$$
The cuts of the function $\phi_N(-\omega,k)$ lie on the negative
semi-axis symmetrically with respect to the cuts in (\ref{12}).
Therefore, $\Pi^0_N(\omega,k)$ has four cuts in the complex
plane of $\omega$, which are shown in Fig.~2. When $\omega>0$,
the cuts are caused by the singularities of $\phi_N(\omega,k)$,
while at $\omega<0$ the cuts are related to the logarithms of
$\phi_N(-\omega,k)$.

The imaginary factors in denominators in  (\ref{9a}) determine the
physical sheet in the complex plane of frequency. On the physical sheet
the first logarithm in (\ref{10}) has the imaginary
part on the upper edge of the cut $I$ equal to $-\pi i$, and  $+\pi i$
on the lower edge of the cut.
In the point $\omega m-kp_F +\frac12k^2$=0, where the cuts
$I$ and $II$ touch each other, the imaginary part of
$\phi_N(\omega,k)$ is continuous but there is a jump of the
derivative. This is shown in \cite{FW}, Fig.~12.9.

Let us now determine the unphysical sheets which we pass on when go
under the cuts $I$ and $I'$ (Fig.~2). Consider the cut $I$.
We denote the first logarithm in (\ref{10}) as
$ln(z_1)\equiv ln\left(\frac{\omega m}
{\omega m-kp_F +\frac12k^2}\right)$.
While the frequency $\omega$ goes along the cut $I$ (\ref{12}):
$\omega=(0,\frac{kp_F}{m} -\frac{k^2}{2m})$ then $z_1$ is
changed in the interval $z_1=(0,-\infty)$. The cut $I$ in the
complex plane of $\omega$ corresponds to the  cut along the
negative real axis in the complex plane of $z_1$. When
we go to the cut $I$ from above (the arrow in Fig.~2a),  this
corresponds that we go to the cut from below in the complex plane
of $z_1$. Going over under the cut,
 we pass on to the unphysical sheet neighboring  with
the physical one on the Riemann surface of $ln(z_1)$. The
values of $ln(z_1)$ on the neighboring sheet differ by
of $(-2\pi i)$ from the values on the physical one at the same
$\omega$.
As example, the value of
$ln\left(\frac{\omega m}{\omega m-kp_F +\frac12k^2}\right)$ on the
lower edge of cut $I$ is
$ln\left(-\frac{\omega m}{\omega m-kp_F+\frac12k^2}\right) +\pi i$.
But the value at the same point  of
$\omega$ on the unphysical sheet is $ln\left(-\frac{\omega
m}{\omega m-kp_F +\frac12k^2}\right)-\pi i$.  Thus, we have the
continuous changing of logarithm along the arrow in Fig.~2a. Then
we conclude that $\Pi(\omega,k)$ in  Eq.~(\ref{1}) changes
continuously as well.
Thus, going over under the cut $I$ in Fig.~2a we pass on to the
unphysical sheet of the complex plane of $\omega$  that is
placed under the physical sheet. Let us denote the unphysical
sheet with the same letter as a cut: $I$.
The part of the unphysical sheet $I$ is
shown in Fig.~2b by the shading with the right  slope.

Now we define an unphysical sheet related to the cut $I'$ (denote
the sheet by $I'$). The cut $I'$ stems from the first logarithm
in $\phi(-\omega,k)$. We designate
$ln(z_1')\equiv
-ln\left(\frac{-\omega m}
{-\omega m-kp_F +\frac12k^2}\right)$ = $
ln\left(\frac{\omega
m+kp_F -\frac12k^2} {\omega m}\right)$.
Going over under the cut of $ln(z_1')$ we add the shift $-2\pi
i$ to the value of $ln(z_1')$. Then, the  logarithm
on the sheet $I'$  under the lower edge of the cut $I'$ is equal to
$ln\left(-\frac{\omega
m+kp_F -\frac12k^2} {\omega m}\right)+\pi i-2\pi i$.  This
expression is used for the analytical continuation to the sheet
$I'$. The part of the sheet $I'$ is shown on the Fig.~2b by the
shading with the left slope.

In the paper we consider the solutions of (\ref{1}) which are placed on
the shaded  parts of the unphysical sheets $I$ and $I'$ in
Fig.~2b. Each logarithmic term in (\ref{9}), (\ref{10}) has its
own Riemann surface.  An important point here is that moving on
Riemann surface of one of logarithms we stand on the
physical sheet for the other logarithms.

Thus, we have built up two unphysical sheets $I$ and $I'$. Let
us show that the values of $\Pi^0_N(\omega,k)$ are real and coincide
on the negative imaginary axes of the sheets $I$ and $I'$.

 On the imaginary axis we denote  $\omega=i\omega_i$.
On the negative imaginary axis of the sheet $I$, the function
 $\Phi_N$ (\ref{9}), (\ref{10}) takes a form
\begin{equation}\label{71}
\Phi_N(i\omega_i,k) = \frac {mp_F}{4\pi^2}\left[1 -
\frac{i\omega_i m}{kp_F} \left(\ln\frac{-i\omega_i m}{i\omega_i
m-kp_F +\frac12 k^2} - \pi i\right) \right.
\nonumber
\end{equation}
$$ \left.
-\frac{i\omega_i m}{kp_F} \left(\ln\frac{i\omega_i m+kp_F
-\frac12 k^2}{i\omega_i m}\right) + \ ... \right].
$$
\begin{equation}\label{81}
= \frac {mp_F}{4\pi^2}
\left[1 +2\frac{\omega_i m}{kp_F}
    \left(arctg\frac{\omega_im}{kp_F-\frac12 k^2} - \frac\pi 2\right)
+  ...\right] \ .
\end{equation}
Here only the first and the second terms of $\phi(\omega,k)$ and
$\phi(-\omega,k)$ (\ref{10}) are shown.

On the other hand, on the  negative imaginary axis of the sheets $I'$
the function $\Phi_N$ (\ref{9}) is
\begin{equation}\label{81a}
\Phi_N(i\omega_i,k) = \frac{mp_F}{4\pi^2}
\left[1 -\frac{i\omega_i m}{kp_F}
\ln\frac{i\omega_i m}{i\omega_i m-kp_F +\frac12 k^2}
\right.
\nonumber
\end{equation}
$$ \left.
-\frac{i\omega_i m}{kp_F}
\left(\ln\frac{i\omega_i m+kp_F -\frac12 k^2}{-i\omega_i m}
    - \pi i\right) +  ... \right]
$$
and after reduction we have the expression (\ref{81}) again.
This is the illustration of the following connection between  the
values of $\Pi^0_N(\omega,k)$ on  $I$ and $I'$  sheets:
\begin{equation}\label{6a}
\left(\Pi^{0}_N(\omega,k)\right)^*_I =
\left(\Pi^{0}_N(-\omega^*,k)\right)_{I'} .
\end{equation}

Thus, $\Pi^0_N(\omega,k)$  is defined on the lower semiplanes of the
unphysical sheets $I$ and $I'$ (the shaded regions in Fig.~2b),
and the values of $\Pi^0_N(\omega,k)$ coincide on the negative imaginary
axes of the sheets $I$ and $I'$.  The special solutions of (1)
which we are interested in, are placed on the shaded part of the
complex plane of $\omega$ in Fig.~2b and on the imaginary axes of
the physical and unphysical sheets.

\section{Investigation of instability at $G'_{NN} < G'_{tr}$}

In this part we consider solutions of the pion dispersion equation (1)
at $G'_{NN} < G'_{tr}$. The definition of $G'_{tr}$ will be given in
the Sect.~5, since it is determined by the behavior of
branches  which are studied in
Sects.~3,4. We start by considering the zero-sound dispersion
equation, and investigate its  solutions for different values of
$G'_{NN}$. We demonstrate the modification of these solutions when
we pass to the pion dispersion equation.

\subsection{Zero-sound dispersion equation}

In this Subsection we consider the unstable
solutions $\omega_{sd}(k)$ of the
zero-sound dispersion equation. For $G'_{NN}<-1$,
they are amplified, placed on the physical sheet and
reflect the instability of nuclear matter in respect to the
small fluctuations of density. For $-1<G'<0$, the branches
$\omega_{sd}(k)$ are damped, and come to  the  negative
imaginary axes (which belongs to the unphysical sheets $I$ and
$I'$).

In this Section we do not take into account the excitations
of isobar and restrict ourselves by
$\Pi(\omega,k)=\Pi_P(\omega,k)$ in (\ref{2f}). The
zero-sound dispersion equation stems from (\ref{7}) and has the
form
\begin{equation}\label{17x}
 -4C_0\Phi_N(\omega,k)=\frac1{G'} .
\end{equation}
The results obtained for this equation are valid both in
spin-isospin and in isoscalar, spin, isoscalar channels.

First we obtain the solutions of Eq.~(\ref{17x}) in the
kinetic theory of Landau  and compare them with the known
results. The function $\Phi_N(\omega,k)$ (\ref{9},\ref{10}) in
the long wavelenths limit is
\begin{equation}\label{13a}
\Phi_N(\omega,k) = 2 \frac{m p_F}{4\pi^2}\left(1-
\frac{\omega m}{2 k p_F}
\ln \frac{\frac{\omega
m}{ k p_F} +1}{\frac{\omega m}{ k p_F}-1}
\right) .
\end{equation}
Substituting this expression in (\ref{17x}), we
obtain the known zero-sound dispersion equation  \cite{LaLi}
\begin{equation}\label{5x}
1+\frac 1{G'} = \frac{\omega m}{2kp_F}ln\left(\frac
{\omega m + kp_F}{\omega m - kp_F}\right) =
\frac{s}{2}\ln\frac{s+1}{s-1} .
\end{equation}
Here we denote $s=\frac{\omega m}{kp_F}$.
When $G'<-1$, this equation has two (positive and negative) solutions
on the imaginary axis of the physical sheet
\cite{LaLi,PN}.  Let us denote $s=i\gamma$, then
Eq.~(\ref{5x}) transforms into
\begin{equation}\label{6x}
1+\frac 1{G'} = \gamma\ arctg\frac 1{\gamma}\ .
\end{equation}

To obtain solutions for $-1<G'<0$, we take   the long
wavelength limit of Eq.~(\ref{81}) and substitute it in (\ref{17x}). On
the negative imaginary axis of the unphysical sheet,
 we get an equation
\begin{equation}\label{9x}
1 + \frac 1{G'} = -\gamma(arctg(\gamma) - \frac {\pi}{2}) .
\end{equation}
 Solutions of this equation are negative, i.e., $\gamma<0$.
This equation coincides with that presented in paper \cite{CC}.
It was pointed out in \cite{CC} that the solution of
Eq.~(\ref{9x}) `` do not correspond to an actual solution". In
agreement with this statement we see that solutions of (\ref{9x})
are located on the negative imaginary  axis belonging both $I$ and
$I'$ unphysical sheets.  It was explained in \cite{CC} the damping of
this mode is the Landau damping \cite{LaLi}.

In Fig.~3 the solutions $\gamma$ to Eq.~(\ref{17x}) are shown at
$G'<0$. They are similar to the solutions at $G'<0$ in Fig.~1 of
 \cite{CC} (except the curve $ab$  at $-1<G'<0$). The shaded
region means that the solution are on the unphysical sheets. In
Fig.~3 it is demonstrated that when the attraction between the
particles is increasing, the damping solution placed on the
unphysical sheet goes over to the physical sheet, turning to
amplified solution.

Considering  the unphysical sheets we can find two solutions of
(\ref{17x}) for any value of $G'<0$ (Fig.~3). However, at
$-1<G'<0$ they are on the different unphysical sheets.  It was
shown that the solutions of Eq.~(\ref{9x}) are negative thus
being located on the unphysical sheet $I$.  Now we  look for
solutions with $\gamma>0$ (curve $ab$ in Fig.~3). The situation
is the same as we had with the sheets $I$ and $I'$. There are
two new unphysical sheets $\tilde I$ and $\tilde I'$ which
overlay the physical one. In Appendix B  the construction of
these sheets is presented. The dispersion equation (\ref{17x})
on the imaginary positive unphysical axis is the same on the
both $\tilde I$ and $\tilde I'$ sheets and has a form

\begin{equation}\label{13}
1 + \frac 1{G'} = -\gamma(arctg(\gamma) + \frac {\pi}{2}) .
\end{equation}
The solutions of this equation are positive. This equation turns
into Eq.~(\ref{9x}) if to change the sign of $\gamma$. The solutions of
this equation are shown by the curve $ab$ in Fig.~3. They are located on
the positive imaginary unphysical  axis
 (this is marked by the  horizontal shading).

Let us compare Fig.~3 and Fig.~1 in \cite{PR}. In paper \cite{PR} only
one curve is presented which corresponds to decreasing of $\gamma$ with
increasing of $G'$. This curve describes  the physical processes: the
amplifying of the amplitudes of the small fluctuations at $G'<-1$ and
the Landau damping of excitations at $-1<G'<0$. According to
\cite{Mi,BaW} in the spin-isospin channel this curve can be considered
as describing the $\pi^+$-type excitations. Another curve shown in
Fig.~3 corresponds to the $\pi^-$-type excitations. Their frequencies
are to be taken with the opposite sign in respect to the ones shown in
Fig.~3.  There are two solutions  for $\pi^0$-type excitations.  One of
them corresponds to the growing excitations (at $G'<-1$) and the second
should be taken with the opposite sign and becomes to degenerate with
the first one in the symmetric and asymmetric matter.

It is well known that the kinetic theory of Landau gives
the linear dependence of
frequencies of  collective excitations  on the  wave
vectors, i.e., $\omega\sim k$. When we pass
to RPA using Eqs.~(\ref{9}), (\ref{10}), the shape of
the dependence changes and we obtain the branches of solutions
$\omega_i(k)$.  In Fig.~4 we show the imaginary branches
$\omega_{sd}(k)$ obtained in RPA.  The shaded fields on the
edges of figures indicate that the branches are located on the
unphysical imaginary axis. We see that the branches change
continuously with $G'$.  When  $-1 < G' < 0$, the branches
$\omega_{sd}(k)$ are located on the unphysical sheet only, but for $G'
< -1$  they go over to the physical sheet in the interval of wave
vectors $k=(0,k_{fin})$.

In analogy with Fig.~3, we can find the second solution of
Eq.~(\ref{1}) for each values of $G'$ and $k$. The ``symmetric"
branch has the conjugated values on the physical sheet
$\omega_{sd}^*(k)$, but it goes to the unphysical sheet $\tilde
I$ when $k>k_{fin}$.

All solutions belong to the same family $\omega_{sd}$. This
family exists at $G'<0$.

\subsection{Pion dispersion equation}
We consider the influence of the pion-nucleon
interaction on the branches $\omega_{sd}(k)$. Following \cite{Mi}
we  include only the direct term of the interaction of the pion with a
particle-hole pair, and do not take into account the noncentral
interaction of the quasiparticles. In this Section we take into account
the isobar-hole pairs contribution to the pion self-energy,
Eq.~(\ref{7}).

First let us investigate how the contribution of isobar-hole
excitations changes the stability conditions. Let us assume
in Eq.~(\ref{17x}) that $\omega$=0 and $k\to 0$. In this limit
this equation reduces to $-1=\frac1{G'_{NN}}$.  This gives the
value of $G'_{NN}$ when instability occurs in the long
wavelength limit.  If we take into account the isobar-hole
loops, this condition takes the form
 \begin{equation}\label{17xy}
 -4C_0\Phi_N(\omega=0,k)\approx
\frac{1+ C_0\frac{16}9\Phi_\Delta(0,k) G^{'2}_{\Delta\Delta}}
{G'_{NN}(1+C_0\frac{16}9\Phi_\Delta(0,k) G'_{\Delta\Delta})
- C_0\frac{16}9\Phi_\Delta(0,k) G^{'2}_{N\Delta}}.
\end{equation}

At $\omega$=0 and $k\rightarrow 0$, $\Phi_\Delta(0,k)$ is
 proportional to nuclear density: $\Phi_\Delta(0,k)\approx
\frac{\rho}{2(m_\Delta-m)}$. If $\rho \rightarrow 0$, the
 corrections produced by $\Phi_{\Delta}$ disappear. As a result
the values of  $\rho_c$ and $k_c$ on the solid lines in Fig.~1
approach the dashed ones. In Fig.~1 the solid lines  are obtained
for the minimal densities when solutions with $\omega$=0 emerge
for every $G'_{NN}$.

It is shown in Fig.~5a, how   $\omega^\pi_{sd}(k)$
are modified when $f_{\pi NN}$ is changed  from 0 to 1. At
$f_{\pi NN}$=0 the branch of solutions of the zero-sound
dispersion equation, Eq.~(\ref{7}), $\omega_{sd}(k)$  is
presented. (It should be noted that the results in  Fig.~5a are
obtained for $\Pi_\Delta(\omega,k)$ contribution being  included
in pion self-energy (\ref{2f}) while those in Fig.~4 do not
include $\Pi_\Delta(\omega,k)$.) As we see from Fig.~5a, the
growth of $f_{\pi NN}$ results in emerging of the amplified
solutions on the physical sheet at $-1< G'_{NN}< G'_{tr}$.

In Fig.~5b the branches $\omega^\pi_{sd}(k)$ are shown for different
values of density of nuclear matter. The curve 3 is calculated for
$\rho=\rho_c$ and  contains the point
$\omega^\pi_{sd}(k_{c})$=0.  It is demonstrated how the zero
value $\omega$=0 is placed on this branch (the point $A$).  At
the point $A$ the condition
$\left(\frac{d\omega^\pi_{sd}(k)}{dk}\right)_{k=k_c}$=0 is
satisfied.  The density $\rho_{c}$ and wave vector $k_{c}$,
corresponding to  zero solution, are presented in Fig.~1 (denoted by
$A$).  When density is lower than the critical one, $\omega^\pi_{sd}(k)$
are completely situated on the unphysical sheet (curve 4) and the
nuclear matter is stable. When $\rho>\rho_c$, the amplified solutions
appear on the physical sheet in the certain interval of $k$ (the curves 1
and 2).

Figures~5a,~b demonstrate the typical behavior of the branches
of solutions of (1) at $G'_{NN} <G'_{tr}$. Thus, we obtain the
instability of nuclear matter related to $\omega^\pi_{sd}(k)$
for all $G'_{NN} <G'_{tr}$. Instability occurs at
$\rho\geq\rho_c$.  The corresponded solutions of the zero-sound
dispersion equation, $\omega_{sd}(k)$, demonstrate instability
at $G'_{NN} < -1$ only.

We refer $\omega^\pi_{sd}(k)$ to $\omega_{sd}$ family.

\section{Investigation of instability at $G'_{NN} > G'_{tr}$}

Now let us consider in Fig.~1 another well studied region
$G'_{NN}\approx 2$. This region of $G'_{NN}$ was investigated in
connection with prediction of the pion condensation.

In Fig.~6 the solutions of Eq.~(1) are presented  at $G'_{NN}=2$
 and $\frac{\rho}{\rho_0}$=1.78. Such figures were published in
papers \cite{Mi}. In Fig.~6 we see the pion branch of solutions
$\omega_\pi(k)$, it starts at the frequency
$\frac{\omega_\pi(k=0)}{m_\pi}=1$. It turns into the free pion
solution as $f_{\pi NN}\rightarrow 0$.  The isobar branch
$\omega_\Delta(k)$ starts at the frequency which is determined
by the isobar and nucleon mass difference. The zero-sound branch
$\omega^\pi_s(k)$ starts at $\omega_s(k=0)$=0, and at certain
value of $k$ it acquires an imaginary part due to  Landau
damping \cite{LaLi}. It goes over  to unphysical sheet related
to the cut $II$, Eq.~(\ref{12}), \cite{DRS,SR}.  Besides the
numerated branches, one more imaginary branch is presented in
Fig.~6. Let us denote it $\omega_c(k)$.  Tracing of this branch
of solutions stimulated  development of the theory of the ``pion
condensation" \cite{Mi}.

In this Section we investigate the family of solutions
$\omega_c$ and show that this family is responsible for the
instability at $G'_{NN} > G'_{tr}$. In Fig.~6  a part of a
branch $\omega_c(k)$ is presented that is located on the
physical sheet in the interval  $k_1 \leq k\leq k_2$. The rest
of this branch is on the unphysical sheet $I$.

In Fig.~7 we present the  same branch $\omega_c(k)$  as in
Fig.~6, but here we show it from another point of view,
i.e., considering the complex plane of $\omega$ (the solid line).
The branch $\omega_c(k)$ starts at $k$=0 at the point `1' and
has the real value $\frac{\omega_c(k=0)}{m_\pi} = 1$. With the
increasing of $k$ it moves to the unphysical sheet $I$. It
follows from (\ref{6a}) that there is another branch of
solutions $\omega'_c(k)$, which starts at the point
$\frac{\omega'_c(k=0)}{m_\pi}=-1$ and is placed on the
unphysical sheet $I'$ (the dashed curve).

Below in paper, the hatching on the edge of figures with the
right slope means that the drawing branches are located on the
unphysical sheet $I$. The left slope means that they are on the
sheet $I'$. The physical sheet is above the axis of abscissas.

In point `2' $\omega_c(k)$ and $\omega'_c(k)$ coalesce on
the imaginary axis. Then $\omega_c(k)$ ascends  the imaginary
axis and $\omega'_c(k)$ descends.
(To determine which of branches go up we can
  take into account the width of isobar in Eq.~(\ref{4f}) as in
\cite{DRS}.) Thus, two simple roots of Eq.~(1) which move on the
unphysical sheets $I$ and $I'$ (when $k$  increase) coalesce on the
imaginary axis at $k=k_m$. Here the roots become imaginary ones,
and continue to move up and down the imaginary axis which belongs
to  both $I$ and $I'$  sheets.

In Fig.~7 the values of $\omega_c(k)$ and $\omega'_c(k)$ are shifted
from the imaginary axis by hand to demonstrate the trajectories of
solutions with increasing of $k$ (it is shown by arrows). The wave
vector $k=k_t$  corresponds to the turning point of
the trajectories `3'. At this point, $\omega_c(k)$ and $\omega'_c(k)$
reach their  maximal and  minimal values. When $k >k_t$, the
branches return along the imaginary axis.  In Fig.~7 the
solutions are presented for $k\leq 2.45 m_\pi$. At larger $k$,
the behavior of branches has a nontrivial character but it is
not related to physical processes and is not discussed here. In
Figs.~6,~7 it is shown how 1)~the branch $\omega_c(k)$ goes
from the unphysical  to the physical sheet at $k=k_1$; 2)~it
becomes imaginary, with growing value of $Im(\omega_c(k)) > 0$,
on the physical sheet; 3)~the branch returns to the unphysical
sheet at $k=k_2$.

In Fig.~8  the complex plane of $\omega$ is presented where
$\omega_c(k)$ and $\omega'_c(k)$ are shown for different densities of
nuclear matter. In these figures (except Fig.~8a), the
branches are drawn up to the turning point on the imaginary
axis, i.e., at $k=(0,k_t)$ (see Fig.~7).  In Fig.~8a the density
is small and the branches do not attain the imaginary axis. In
Fig.~8b the branches encounter on the imaginary axis but
completely stay on the unphysical sheets ($k_t=1.78m_\pi$). In
Fig.~8c  the result for the critical density $\rho_c$  is
presented. On the real axis the solution with $\omega$=0
appears. Here $k_t=k_c$ ($k_t=1.81m_\pi$)  and we see that the
conditions $\omega_c(k_c)$=0 and
$\left(\frac{d\omega_c(k)}{dk}\right)_{k=k_c}$=0 are satisfied.
The wave vector and density corresponding to the point $B$ in Fig.~8c
are marked by $B$ in Fig.~1. In Fig.~8d it is demonstrated that at
$\rho>\rho_c$ the solutions with $Im(\omega_c(k)) > 0$
($k_t=1.88m_\pi$) exist on the physical sheet.

The critical densities and wave vectors in Fig.~1 for $G'_{NN} >G'_{tr}$
correspond to the solutions denoted by $B$ in Fig.~8c. Such solutions
$\omega_c(k)$ can be  found not only when $G'_{NN} > G'_{tr}$,
but for $G'_{NN} < G'_{tr}$ also. It should be mentioned, that an
attempt to shift the pion condensation to  larger densities means that
$\omega_c(k)$ go on  to stay on the unphysical sheet at
larger densities. There are several reasons that can influence our
results. It was shown in \cite{DRS} that the important factor shifting
the pion condensation to the larger densities is the decreasing of the
effective nucleon mass with growth of density. This decreasing is
predicted by the scaling of Brown-Rho \cite{BR} and QCD sum rules
\cite{MRL,DRSF}.  However, at smaller values of the nucleon mass
the lowest order expansion in powers of $\frac{p^2}{2m}$ in
Eq.~(\ref{9a}) is not valid any more, thus requiring relativistic
kinematics for the nucleon.  Besides, the formation of isobar Fermi
 surface must be taken into account, as soon as the density becomes so
large that the quasiparticle energy on the Fermi surface exceeds the
difference of the isobar and nucleon  mass.

In Figs.~7,~8 the branches $\omega_c(k)$ start at the point
$\omega_c(k=0)=m_\pi$. Inclusion of the scalar part of the pion
self-energy $\Pi_S$, Eq.~(\ref{1f}),  displaces the
initial value to $\omega^2_c(k=0)=m^2_\pi + \Pi_S$ (Appendix A).
However, this does not modify the curves in Fig.~1
qualitatively.

\section{Determination of $G'_{tr}$}

In Sects.~3, 4 we have presented  zero-frequency solutions
of Eq.~(1), which appeared for different values of $G'$.
 In this Section the ``transitional" value  $G'_{tr}$
is obtained.  In Sects.~3,4 it was shown
that at $G'_{NN} < G'_{tr}$ the zero-frequency solutions belong to
the family $\omega_{sd}$ and  at $G'_{NN} > G'_{tr}$ they
refer to  $\omega_c$. In this Section the
transition of solution with $\omega$=0 from one family  to
another is demonstrated.

In the vicinity of $G'_{tr}$,
at $G'_{NN} \approx G'_{tr}$, the branches belonging to
 $\omega^\pi_{sd}$ and $\omega_c$ families are placed
close to each other on the same unphysical sheets. In Figs.~5,~7,~8
these branches are shown separately. In Fig.~9 the branches of the both
families are shown simultaneously.

In Fig.~9a at $G'_{NN}=-0.0020$ and $\rho=\rho_c$,
the branches $\omega_c(k)$, $\omega'_c(k)$ and
$\omega^\pi_{sd}(k)$ are presented simultaneously.
The critical value of the wave vector is
$k_c=1.61m_\pi$. The branches $\omega_c(k)$ and $\omega'_c(k)$ behave
like  in Fig.~8a. The branch $\omega^\pi_{sd}(k)$ is shown
by the dotted curve and it is presented in the interval  $k=(0, k_c)$.
The behavior of $\omega^\pi_{sd}(k)$ is similar to the curve 3 in
Fig.~5b between $k$=0 and the point $A$, which corresponds to
$k=k_c$.

In Fig.~9a number `1' denotes the start of branches at $k$=0;
`2' means the turning point of
$\omega^\pi_{sd}(k)$:~$k_t$=0.077$m_\pi$; `3' corresponds to
$k$=0.1$m_\pi$. We see that $\omega^\pi_{sd}(k)$ descends when
$k$ changes from '0' to $k_t$ and then ascends and turns to zero
at $k=k_c$:~$\omega^\pi_{sd}(k=k_c)$=0 (point $C$ in Fig.~1).

Now let us enlarge a little the coupling
constant:~$G'_{NN}$=-0.0018.  The following calculations are
made at the critical density $\rho=\rho_c$, corresponding to
this value of $G'_{NN}$. In this case we have a coalescence of
$\omega_c(k)$ and  $\omega'_c(k)$ on the imaginary axis. The
wave vector of coalescence is equal to $k_m$=0.0962$m_\pi$. In
this difficult for drawing case, the presentation of the
branches is made in two steps. In Fig.~9b the branches are shown
at $k\leq 0.094m_\pi$, i.e., before coalescence and in Fig.~9c at
$k\geq 0.094m_\pi$. The position of branches at $k$=0.094$m_\pi$
are marked by `2'.

In Fig.~9b it is shown that the position of  $\omega^\pi_{sd}(k)$ at
point `2' is much lower on the negative imaginary axis than the
$Im\omega_c(k)$ and  $Im\omega'_c(k)$. Further, in Fig.~9c we see the
coalescence of $\omega_c(k)$ and $\omega'_c(k)$ at $k=k_m$ (point `3').
Then $\omega_c(k)$ and $\omega'_c(k)$ go in the opposite directions
along the imaginary axis.

In Fig.~9c it is shown that the branch $\omega'_c(k)$ descends and
meets the branch $\omega^\pi_{sd}(k)$ in the point `4' (at $\tilde
k_m$=0.109$m_\pi$).
[The branch $\omega^\pi_{sd}(k)$ has turned upward at
$k$=0.104$m_\pi$.] After the second  coalescence, the branches
$\omega'_c(k)$ and $\omega^\pi_{sd}(k)$ go in the opposite sides
on the complex plane. Thus, the branch $\omega'_c(k)$ has
blocked $\omega^\pi_{sd}(k)$ on the unphysical sheet.

The branch $\omega_c(k)$ goes over to the real axis and a solution with
 $\omega$=0 (point $D$) belongs to this
branch:~$\omega_c(k_c)$=0 at $k_c$=1.60$m_\pi$ (the point $D$ in
Fig.~1).

Thus, the value of $G'_{tr}$  is determined by the interaction of
branches on the unphysical sheets shown in Fig.~9. We can
conclude that $G'_{tr}$ is a special value of the spin-isospin
effective quasiparticle interaction  when 1)~there is  coalescence of
$\omega_c(k)$ and $\omega'_c(k)$ on the imaginary axis at the critical
density $\rho=\rho_c$  and 2)~$Im\omega^\pi_{sd}(k_m)\leq$
 $Im\omega_c(k_m)$, $Im\omega'_c(k_m)$. The second condition is
satisfied as a rule, since $Im\omega^\pi_{sd}(k_m)$ tends to $-\infty$
as $G'\rightarrow -0$ (compare with Fig.~3).

In Fig.~9 the results  are presented for
$\Pi(\omega,k)=\Pi_N(\omega,k)$ (\ref{1f}),
i.e., without the excitation of isobars. In this case we have obtained
$G'_{tr}$=-0.0019.  If to solve Eq.~(1) with
$\Pi_P(\omega,k) = \Pi_N(\omega,k) + \Pi_\Delta(\omega,k)$,
then $G'_{tr}$=0.0020. The
influence of $S$-wave  pion-nucleon interaction is discussed in
Appendix A.

It is interesting to mention that the figure similar to Fig.~9 is
presented in \cite{PR} for the solutions of the zero-sound dispersion
equation. Their figure demonstrates the
behavior of solutions for the different values of $F_0$ and the
branches of solutions depend on $F_1$ instead of $k$ (this means the
dependence on the value of the quasiparticle effective mass $m^*$ along
the branch).  In paper \cite{PR} $F_0$ and $F_1$ are the zero and the
first multipolar components of the scalar-isoscalar effective
interaction of the quasiparticles.

\section{ Discussion}

\noindent
{\bf A.} In this paper two families of solutions to the pion dispersion
equation are calculated, which determine the  stability boundary
of the nuclear matter  (the critical density $\rho_c$)
depending on the value of spin-isospin coupling constant $G'$.
It is shown that there is  a special value of $G'= G'_{tr}$
such that for $G'< G'_{tr}$ the stability boundary are related
to the family $\omega^\pi_{sd}$ and  for $G'> G'_{tr}$ it
refers to $\omega_c$.

It is interesting to investigate, whether we can  obtain the
information about the difference of the phase transitions  at
$G'<G'_{tr}$ and $G'>G'_{tr}$ using the  branches of solutions.
To do this let us consider the rate of growth of excitations in the
vicinity of $G'_{tr}$. For $\rho > \rho_c$,  Eq.~(1) has
solutions with $Im \omega_i(k) > 0$. In Figs.~5,~7,~8d such
solutions are presented. We determine $\Gamma(k)=-i\omega_i(k)$,
$i=sd,c$.  We can take several values of $G'$ close to
$G'_{tr}$ and calculate the  maximal values of $\Gamma(k)$
(i.e., $\Gamma(k_t)$) depending on $\rho$ for the every $G'$.
In Fig.~10 these results are demonstrated for
$G'=G'_{tr}, G'_{tr}\pm 0.02, G'_{tr}\pm 0.2$, where
$G'_{tr}$=-0.0019.  In this figure we cannot observe the essential
difference in  the dependence of $\Gamma(k_t)$ on $\rho$ for
$G'<G'_{tr}$ and $G'>G'_{tr}$. It is possible to conclude that either
this test is not sensitive to the type of excitations, or the character
of the phase transition  changes continuously with $G'$ from long
wavelength instability up to the pion condensation.

\noindent
{\bf B.} Now let us consider the singularity of the integrals
containing the pion propagator $D$ (1): $F=\int
D(\omega,k)f(\omega,k)d\omega d^3k$, in the poles of $D(\omega,k)$ when
$\rho=\rho_c$ and $k\to k_c$.  The function $f(\omega,k)$ is not
related to the poles of $D(\omega,k)$. In the Sect.~3.1 for the
zero-sound dispersion equation (\ref{7}), it was shown that
there are two solutions on the physical sheet. At $-1<G'<0$, the
amplified solutions go over to the unphysical sheets $I$, $I'$
and the second branch goes over to the unphysical sheets $\tilde
I$, $\tilde I'$ (Fig.~3).

Similar situation takes place for the pion dispersion equation. Up to now
we considered the solutions which are located on the physical and
unphysical $I$ and $I'$ sheets, but there are the ``symmetrical"
solutions. They have the conjugated values on the physical sheet and
their unphysical sheets are $\tilde I$ and $\tilde I'$. This means that
at $\rho=\rho_c$ and  $k=k_c$ not only $\omega_i(k_c)$=0
($i=sd,c$) but the ``symmetrical" branch as well. As a result,
the function $F$ has a form (we leave the singular part only) $$
F\approx \int \frac{d\omega
 d^3k}{(\omega^2-\omega^2_i(k))^2}f(\omega=0,k_c).
$$
Since we are interested in such zero solutions that
$\left(\frac{d\omega_i(k)}{dk}\right)_{k=k_c}$=0, then
$\omega_i=a(k-k_c)^2$ as  $k\to k_c$ and the integral diverges along
the curves in Fig.~1. This correspond to situation of the
phase transitions \cite{LaLi}.

\noindent
{\bf C.} In the paper we do not consider the case when
 $G'_{N\Delta}$, $G'_{\Delta\Delta}$ are modified
simultaneously with $G'_{NN}$. But
as it is shown above, the existence of $G'_{tr}$  and
$\omega_c(k)$, $\omega_{sd}(k)$ does not depend on whether we take
excitation of the isobar into account or not. Therefore we do
 not expect the qualitative changes in our results if we change
the values of $G'_{N\Delta}$ and $G'_{\Delta\Delta}$ together with $G'_{NN}$.

\section{ Conclusion}

In the paper we investigate of the zero-frequency solutions of the
pion dispersion equation depending on the values of the spin-isospin
coupling constant of quasiparticles $G'$. For every
$G'$,  the critical density $\rho_c$ and the wave vector
$k_c$  corresponding to the solution with $\omega$=0 were
 calculated.  Then the branches of solutions, containing these
zero solutions, are constructed and it is determined to what
type of excitations (to what family) these  solutions referred.

It is shown that the zero-frequency solutions belong to the
different types of excitations, depending the value of $G'$. In
all models used in calculations, the ``transitional" value
$G'_{tr}$ ($|G'_{tr}|\ll 1$) is found, such that at $G'=G'_{tr}$
the zero solution passes from the branches of the type
$\omega_{sd}$ to the ones of type $\omega_c$.

At $G' < G'_{tr}$, the unstable amplified branches of solutions
for the zero-sound dispersion equation ($\omega_{sd}(k)$) and
for the pion dispersion equation ($\omega^\pi_{sd}(k)$) are
constructed (Figs.~3,~4,~5).
In Fig.~5a we see that if to take the
pion-nucleon coupling constant $f_{\pi NN}$ value to be very small,
$\omega_{sd}(k)$ and $\omega^\pi_{sd}(k)$ are very close. When we
increase $f_{\pi NN}$ up to 1, the behavior of $\omega^\pi_{sd}(k)$ is
change: the damping solutions become amplified ones (Fig.~5). We obtain
that the zero-frequency solutions of Eq.~(1) at $G' < G'_{tr}$ belong
to the same family of solutions which is responsible for the long
wavelength instability in the kinetic theory of Landau.

Figure 1 demonstrates that the stability boundary continuously
extends to $G' > G'_{tr}$. But at these $G'$ the unstable
solutions belong to family  $\omega_c$. It is shown that this family
contents the solutions responsible for the pion condensation.

It is possible to expect that a new state of medium after  phase
transition is different depending the value of $G'$. We investigated
the dependence of the maximal growth rate of perturbations
at several $G'$ in the vicinity of $G'_{tr}$ on density. But this
dependence does not exhibit the noticeable modifications at
$G'_{tr}$.

Author thanks S.~V. Tolokonnikov for the constructive
criticism. Also, author thanks V.~R.~Shaginian and
E.~G.~Drukarev
and especially M.~G.~Ryskin for the numerical fruitful
discussions.
The work was supported by the grants RFBR-03-02-17724.

\section{ Appendix A.}

In this appendix the scalar part of the pion self-energy,
$\Pi_S$, (2) and its influence on the behavior of the curves in
Fig.~1 are considered.  We construct $\Pi_S$ using the relation
of Gell-Mann--Oakes--Renner \cite{DRS}, \cite{GMOR}. It permits
to express the pion mass in medium through the quark condensates
in nuclear matter
$$
 \label{a1} m^{*2}_\pi\ =\ -\frac{\langle NM|\bar
qq|NM\rangle(m_u+m_d)}{ 2f^{*2}_\pi}\ , \eqno{(A1)}
$$
where $m_u,m_d$ - masses of
 $u$- and $d$-quarks ($m_u+m_d$=11~MeV),
$f^*_\pi$ - coupling constant of the pion decay in the medium
(here we assume that $f^*_\pi=f_\pi$=92~MeV, i.e., the vacuum and
medium values of $f_\pi$ coincide).
The value $\kappa=\langle NM|\bar qq|NM\rangle$ is a scalar
quark condensate in nuclear matter. It may be present as
$$
\kappa\ =\ \kappa_0+\rho\langle N|\bar qq|N\rangle +...
\eqno{(A2)}
$$
Here $\kappa_0$ is a scalar quark condensate in vacuum,
$\kappa_0$=-0.03~GeV$^3$; $\langle N|\bar qq|N\rangle$ is the
matrix element of the scalar quark condensate calculated over
the nucleon $\langle N|\bar qq|N\rangle\simeq 8$. Then from (A1)
we obtain
$$
m^{*2}_\pi\ =\ m^2_\pi-\rho\,\frac{\langle N|\bar
qq|N\rangle (m_u+m_d)}{2f^2_\pi}\ .
\eqno{(A3)}
$$
On the other hand, we can define $m^{*2}_\pi$ using Eq.~(1) as
$$
m^{*2}_\pi\ =\ m^2_\pi+\Pi(\omega,k=0)\ .
\eqno{(A4)}
$$

The form of $\Pi_S$ can be obtained comparing (A2) and (A3)
provided that $\Pi_P(\omega,k=0)=0$
$$
\Pi_S\ =\ -\rho\ \frac{\langle N|\bar qq|N\rangle(m_u+m_d)}{
2f^2_\pi}\ .
\eqno{(A5)}
$$

In Fig.~1 the dotted lines correspond to the critical density
$\rho_c$ and wave vector $k_c$ obtained for Eq.~(1) with the
pion self-energy (\ref{1f}). (Recall that
for each $G'$ in Fig.~1 we show the minimal $\rho_c$ when
solution $\omega$=0 appears).  In this model the transition
value of  $G'$ is $G'_{tr}$=0.0026.

\section{Appendix B}
Here we construct the unphysical sheets $\tilde I$ and $\tilde I'$.
The solutions of Eq.~(19) that are presented by  the curve
$ab$ in Fig.~2 are placed on these sheets.

To pass on to the sheet $\tilde I$ we do the analytical continuation
in $\omega$ of
$ln(z)\equiv ln\left(\frac{\omega m}
{\omega m-kp_F +\frac12 k^2}\right)$
from the lower edge of the cut $I$ upwards, adding $2\pi i$ to
$ln (z)$.
The sheet $\tilde I$ is situated above the physical sheet.
The value of $ln\left(\frac{\omega m}
{\omega m-kp_F +\frac12 k^2}\right)$  above the upper edge of  the cut
$I$ is $ln(-z)-\pi i+2\pi i$. This expression is continued on the sheet
$\tilde I$.

In similar way we construct the sheet $\tilde I'$ making the
analytical continuation of
$ln(z')\equiv ln\left(\frac
{\omega m+kp_F -\frac12 k^2}{-\omega m}\right)$
from the lower edge of cut $I'$, adding $2\pi i$ to logarithm
values.

\newpage

\section{Figure captions}
\noindent
FIG.~1.
Critical values of nuclear density $\rho_c$ and  wave
vector $k_c$, at which in Eq.~(1) solutions $\omega$=0 appear.
Results for three models of the pion self-energy are
presented:  solid lines correspond
$\Pi=\Pi_N + \Pi_\Delta = \Pi_P$ (\ref{1f}), (\ref{2f});
dashed curves are  for  $\Pi= \Pi_N$;
 dotted curves are for the total $\Pi= \Pi_P+\Pi_S$
(\ref{1f}).

\noindent
FIG.~2. Complex plane of the frequency  $\omega$.
a) Numbers $I$ and $II$ denote the  cuts (\ref{12})  of
$\phi(\omega,k)$ and $I'$ and $II'$ denote the  cuts of
$\phi(-\omega,k)$;
b) Unphysical sheets are shown:
 sheet $I$ (the shading with the right  slope)  and
 sheet $I'$ (the shading with the left  slope).

\noindent
FIG.~3. Solutions of Eqs.~(\ref{6x}),~(\ref{9x}),~(\ref{13})
at $G'<0$ in the kinetic theory of Landau. The shading with the
left  slope marks unphysical sheet $I$ and the horizontal shading
refers to $\tilde I$.
Results are obtained at  $\rho=\rho_0$.

\noindent
FIG.~4. Imaginary solutions $\omega_{sd}(k)$ to
Eq.~(\ref{17x}) obtained in RPA for  following values of
 $G'$:  (1) -0.4; (2) -0.9; (3) -1.02; (4) -1.1; (5) -1.2.
Results are obtained at  $\rho=\rho_0$.

\noindent
FIG.~5.
Imaginary solutions $\omega_{sd}(k)$ of  Eq.~(\ref{7}). 
a)~Dependence of $\omega_{sd}(k)$  on  $f_{\pi NN}$:
the curve (1) $f_{\pi NN}$=0; (2) 0.4; (3) 0.6; (4) 1.0.
The results are obtained at  $\rho=\rho_0$ and
$G'_{NN}$=-0.4.
b)~Dependence of $\omega_{sd}(k)$ on the value of density: the curve (1)
$\rho/\rho_0$=1; (2) 0.318; (3) 0.265; (4) 0.221.
The results are obtained at  $f_{\pi NN}$=1 and $G'_{NN}$=-0.4.
Point $A$ marks solution with $\omega$=0.

\noindent
FIG.~6. Different types of solutions of Eq.~(1):
spin-isospin zero-sound branch
$\omega^\pi_s(k)$, pion branch $\omega_\pi(k)$, isobar branch
$\omega_\Delta(k)$ and condensate branch $\omega_c(k)$.
Results are obtained at $\frac{\rho}{\rho_0}$=1.78 and $G'_{NN}$=2.
We computed that  $k_1$=1.35$m_\pi$, $k_2$=2.35$m_\pi$.

\noindent
FIG.~7. Branches  $\omega_c(k)$ and $\omega'_c(k)$ on the complex
plane of frequency.
The shading with the right (left) slope marks the unphysical
sheet $I$ ($I'$).
The physical sheet is above the abscissas axis. Axes of
reference are shown by the dotted lines. The solid curve
correspond to $\omega_c(k)$ and the dashed one is for
$\omega'_c(k)$.  The number `1' marks  start of branches
at $k$=0; `2' is the coalescence point at
$k_m$=1.28$m_\pi$; `3' is the turning point at $k_t$=
1.88$m_\pi$.  Variables $k_1, k_2$, $\frac{\rho}{\rho_0}$
and $G'_{NN}$ are the same as in Fig.~6.

\noindent
FIG.~8. Branches  $\omega_c(k)$ and $\omega'_c(k)$ on the
complex plane of frequency at  different densities
$\frac{\rho}{\rho_0}$.  In  figures b,c,d  branches are
shown at $k\leq k_t$. Point $B$ marks  solution with
$\omega$=0.  Notations of lines and sheets are the same as in
Figs.~6,~7.

\noindent
FIG.~9. Transition of a zero solution from  family
$\omega_{sd}$ to  family  $\omega_c$. The solid, dashed
and dotted curves  denote $\omega_c(k)$, $\omega'_c(k)$
and $\omega^\pi_{sd}(k)$, correspondingly.
a) $G'$=-0.0020. The number `1' corresponds to $k$=0; `2' means
 turning point of $\omega^\pi_{sd}(k)$:
$k_t$=0.077$m_\pi$; `3' $k$=0.1$m_\pi$. Arrows show 
movement of  branches with $k$. Point $C$ corresponds to
$\omega^\pi_{sd}(k_c)$=0.
b) $G'$=-0.0018. Branches are shown for $k\leq 0.094m_\pi$ (point
`2').
c) Continuation of figure b. Branches are shown for
$k>0.094m_\pi$.  Point `3' marks the coalescence point of
$\omega_c(k)$ and  $\omega'_c(k)$ at $k$=0.0962$m_\pi$.
Point `4' shows the coalescence point of
$\omega'_c(k)$ and $\omega^\pi_{sd}(k)$  at $k$=0.109$m_\pi$.
Point $D$ corresponds to $\omega_c(k_c)$=0.

\noindent
FIG.~10. Dependence of the maximal growth rate of
fluctuations on density.  The curve (1) corresponds to
$G'=G'_{tr}$=-0.0019; (2) $G'=G'_{tr}$-0.02; (3)
$G'=G'_{tr}$+0.02; (4) $G'=G'_{tr}$-0.2; (5) $G'=G'_{tr}$+0.2.

\newpage

%Fig.1
\begin{figure}
\centering{\epsfig{figure=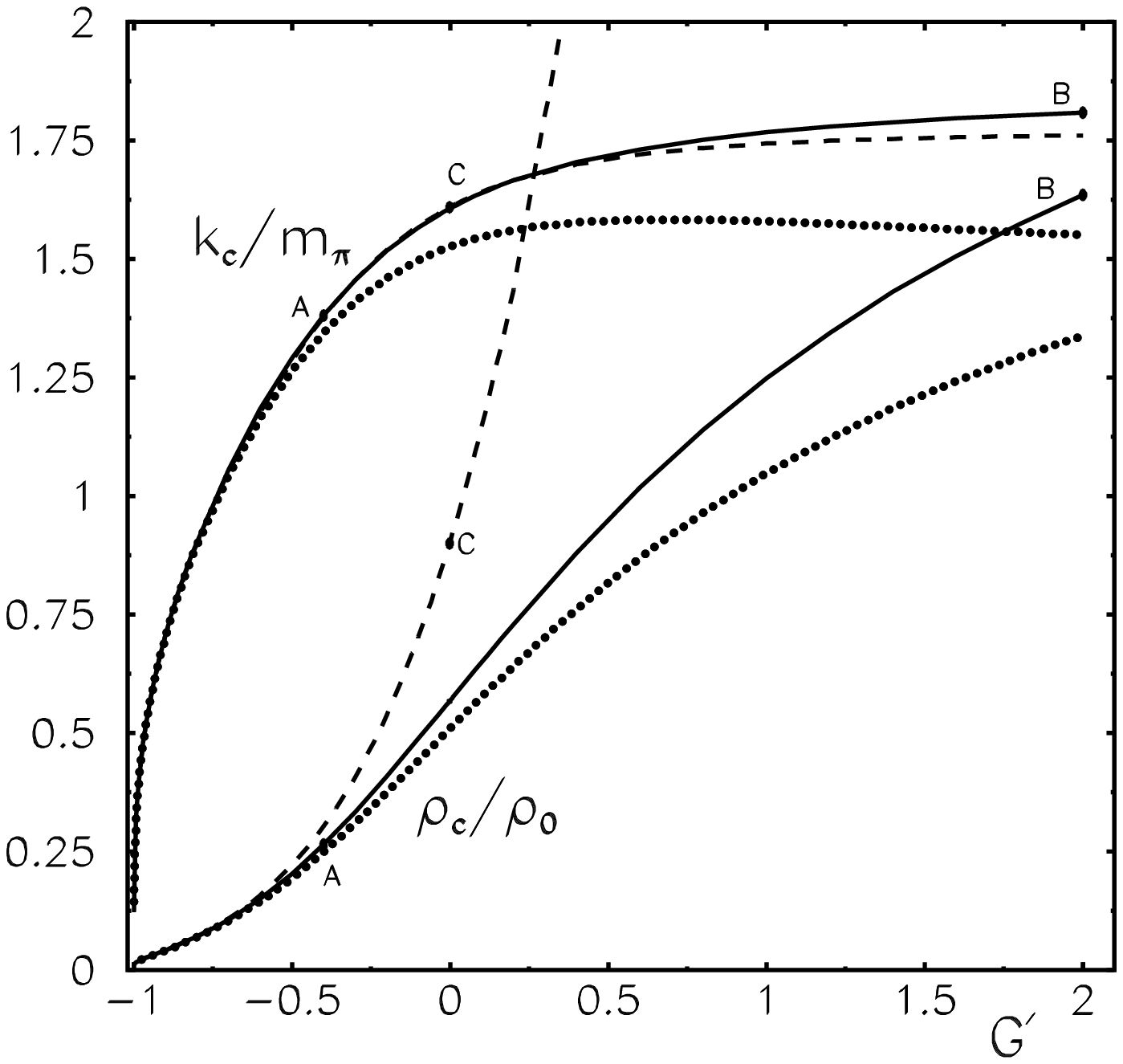,width=8cm}}
\caption{}
\end{figure}

%Fig.2
\begin{figure}
\centering{
\epsfig{figure=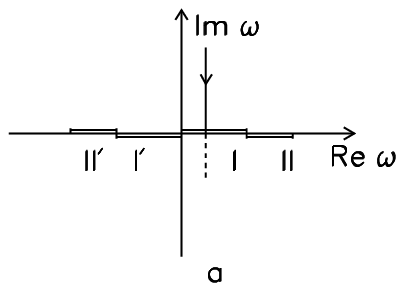,width=8cm}
\epsfig{figure=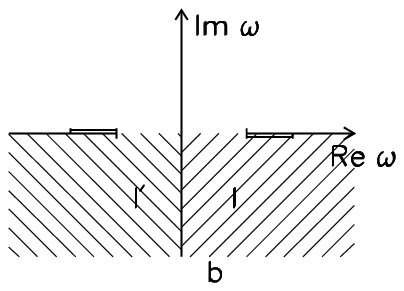,width=8cm}
}
\caption{}
\end{figure}

%Fig.3
\begin{figure}
\centering{\epsfig{figure=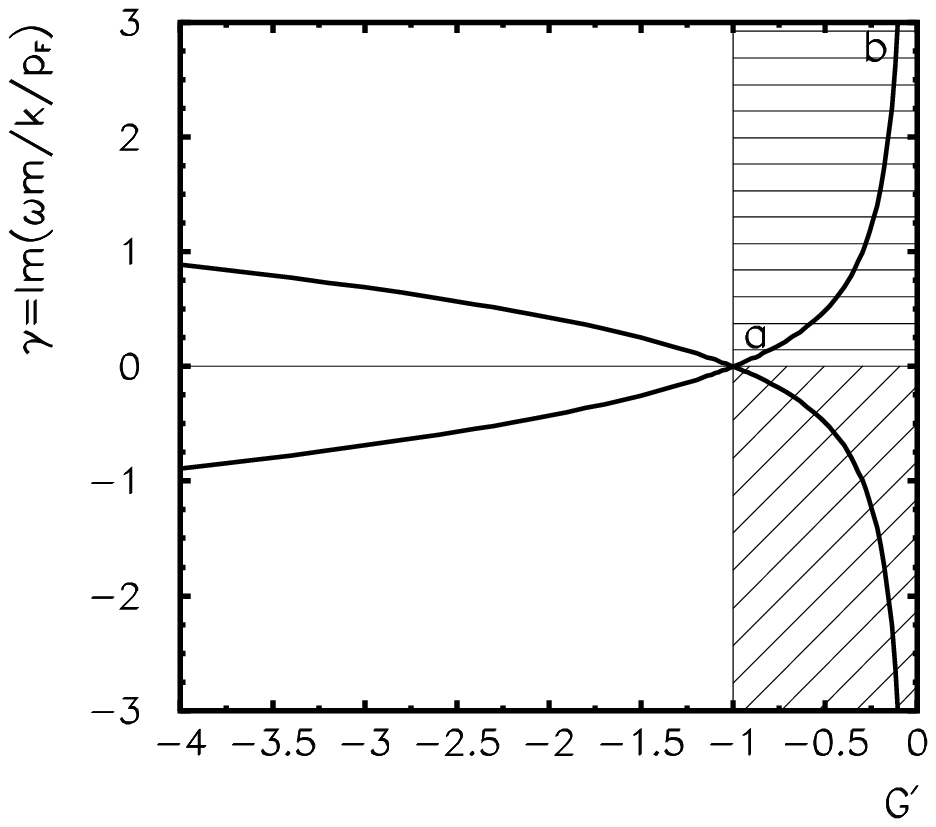,width=8cm}}
\caption{}
\end{figure}

%Fig.4
\begin{figure}
\centering{\epsfig{figure=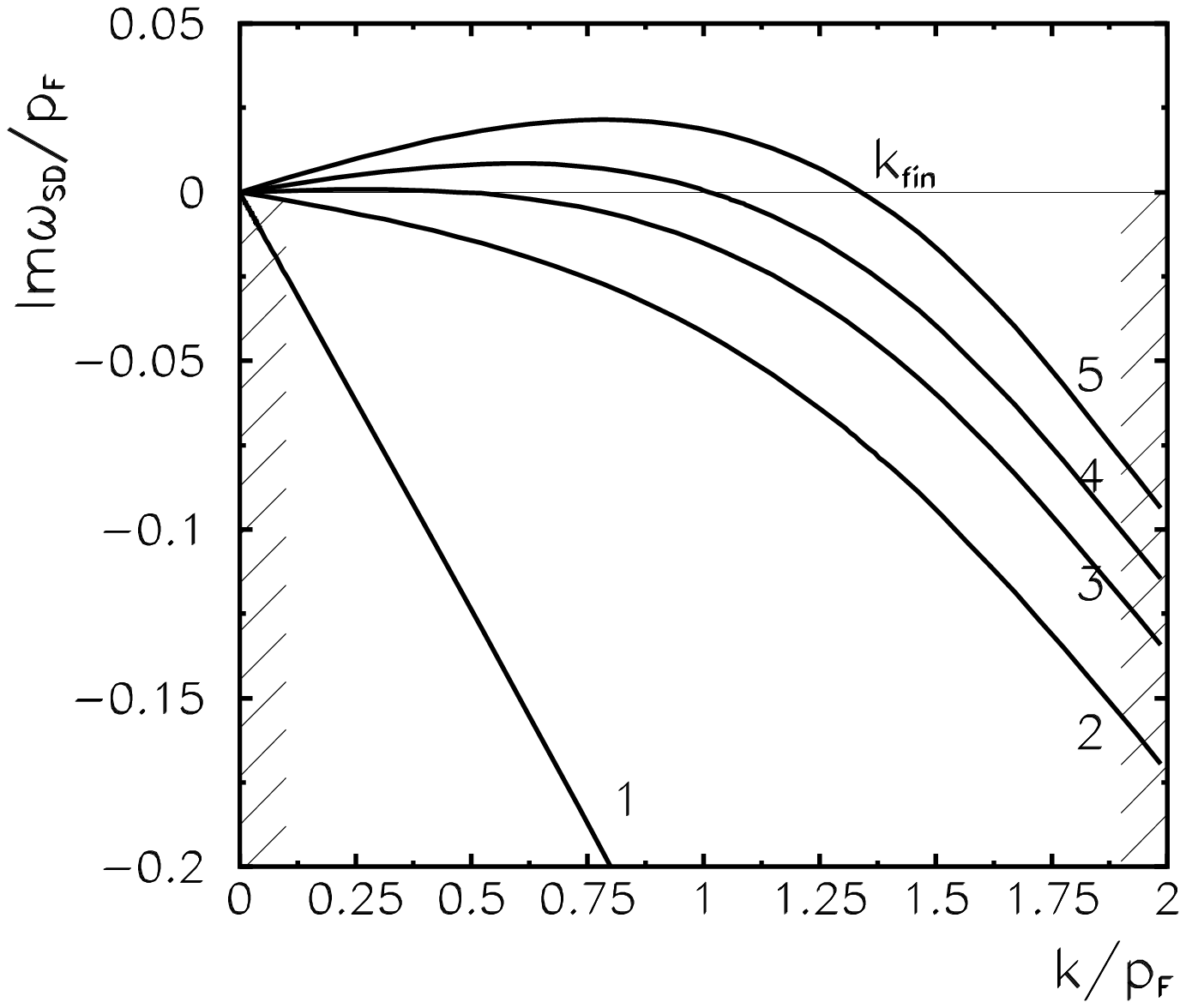,width=9cm}}
\caption{}
\end{figure}

%Fig.5
\begin{figure}
\centering{\epsfig{figure=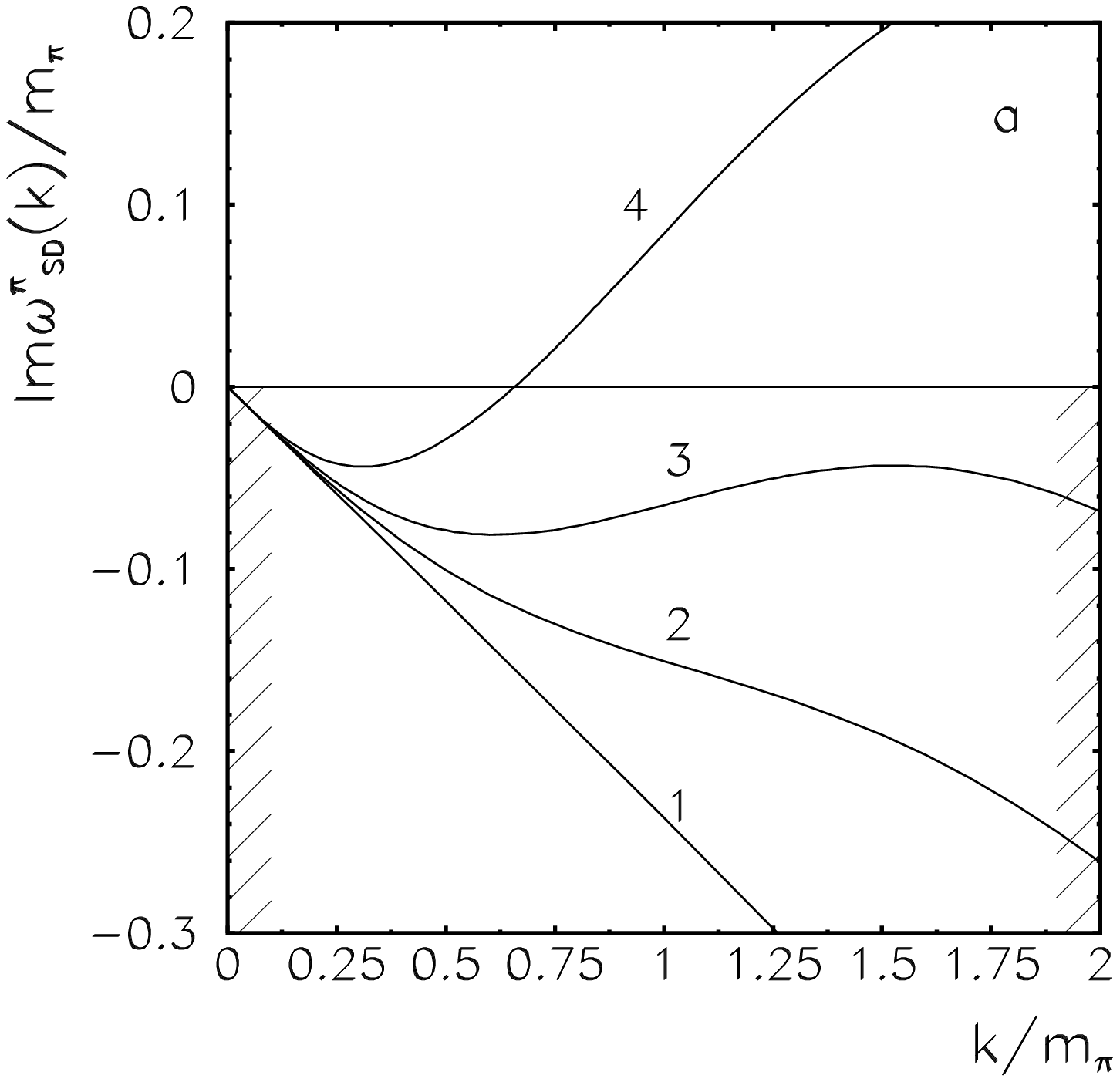,width=8cm}}
\centering{\epsfig{figure=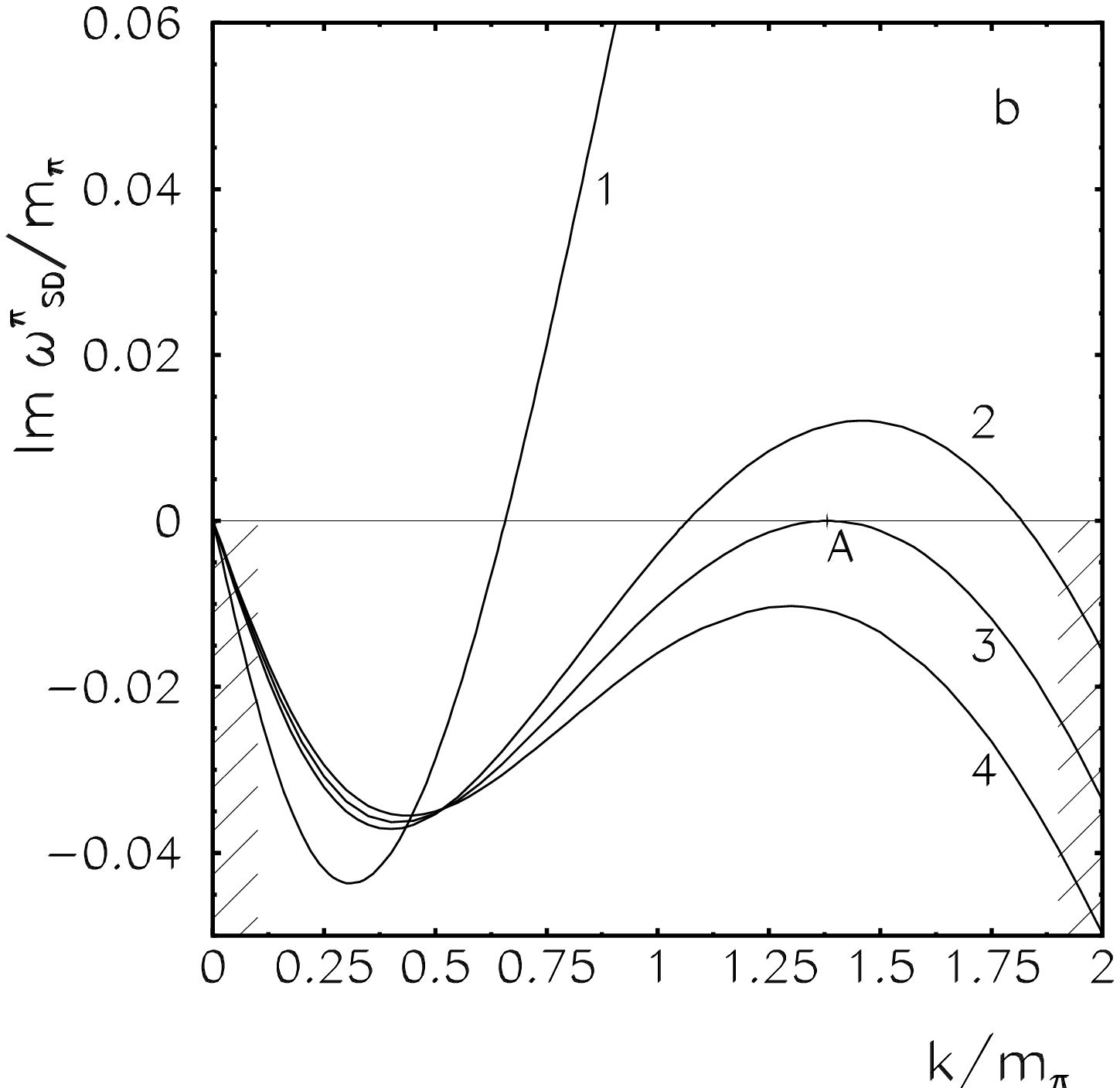,width=8cm}}
\caption{}
\end{figure}

%Fig.6
\begin{figure}
\centering{\epsfig{figure=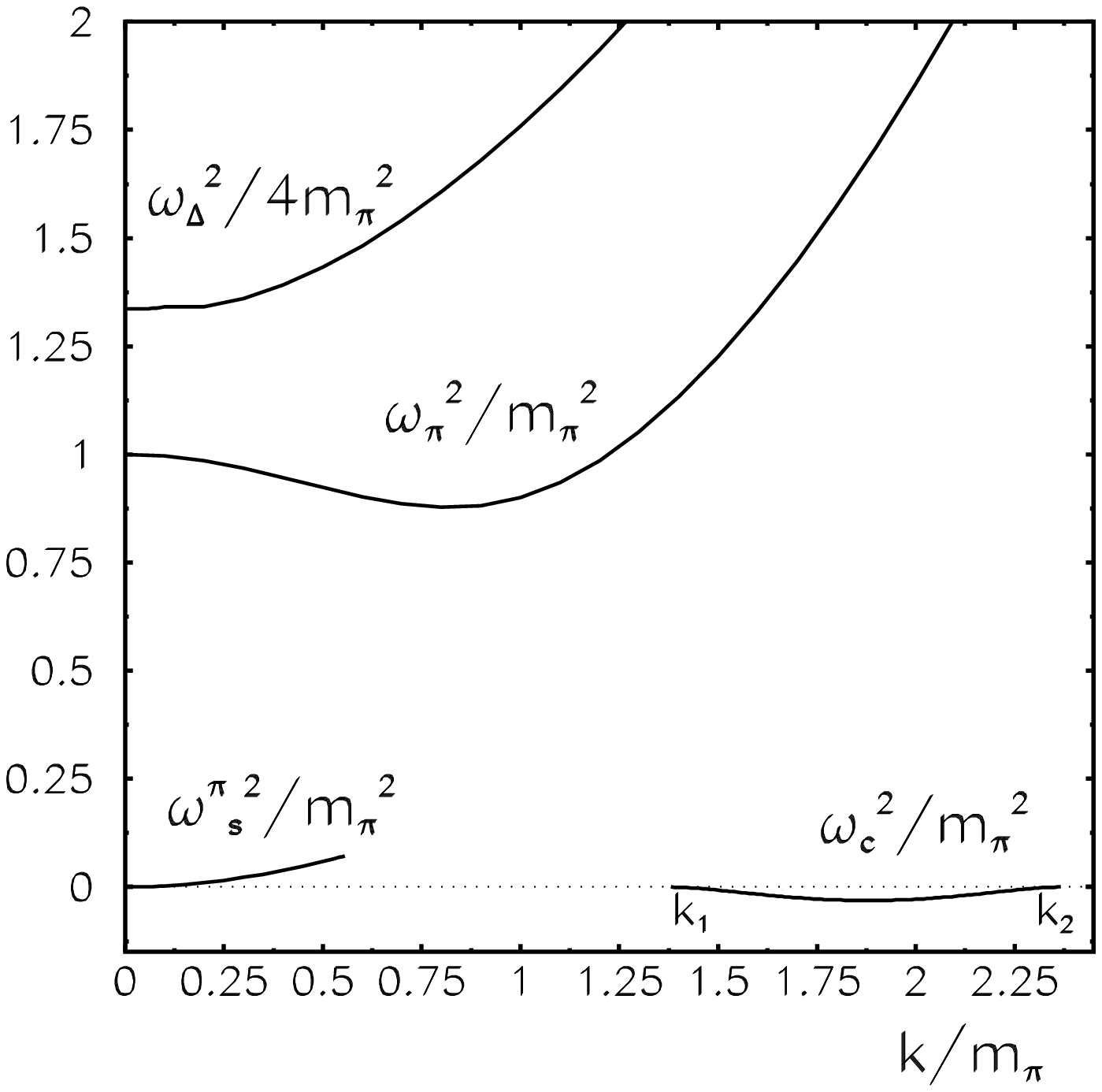,width=8cm}}
\caption{}
\end{figure}

%Fig.7
\begin{figure}
\centering{\epsfig{figure=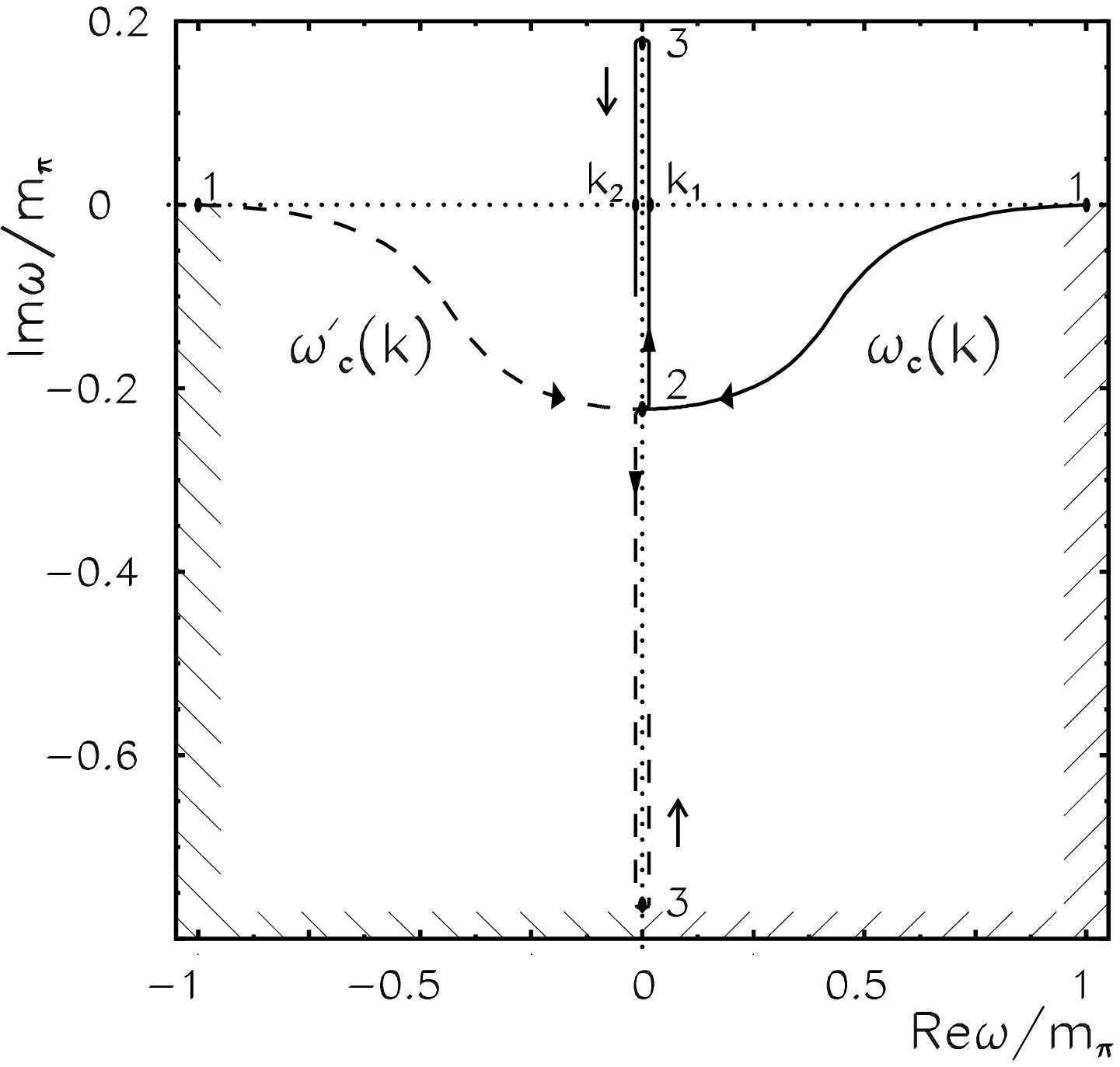,width=9cm}}
\caption{}
\end{figure}

%Fig.8
\begin{figure}
\centering{\epsfig{figure=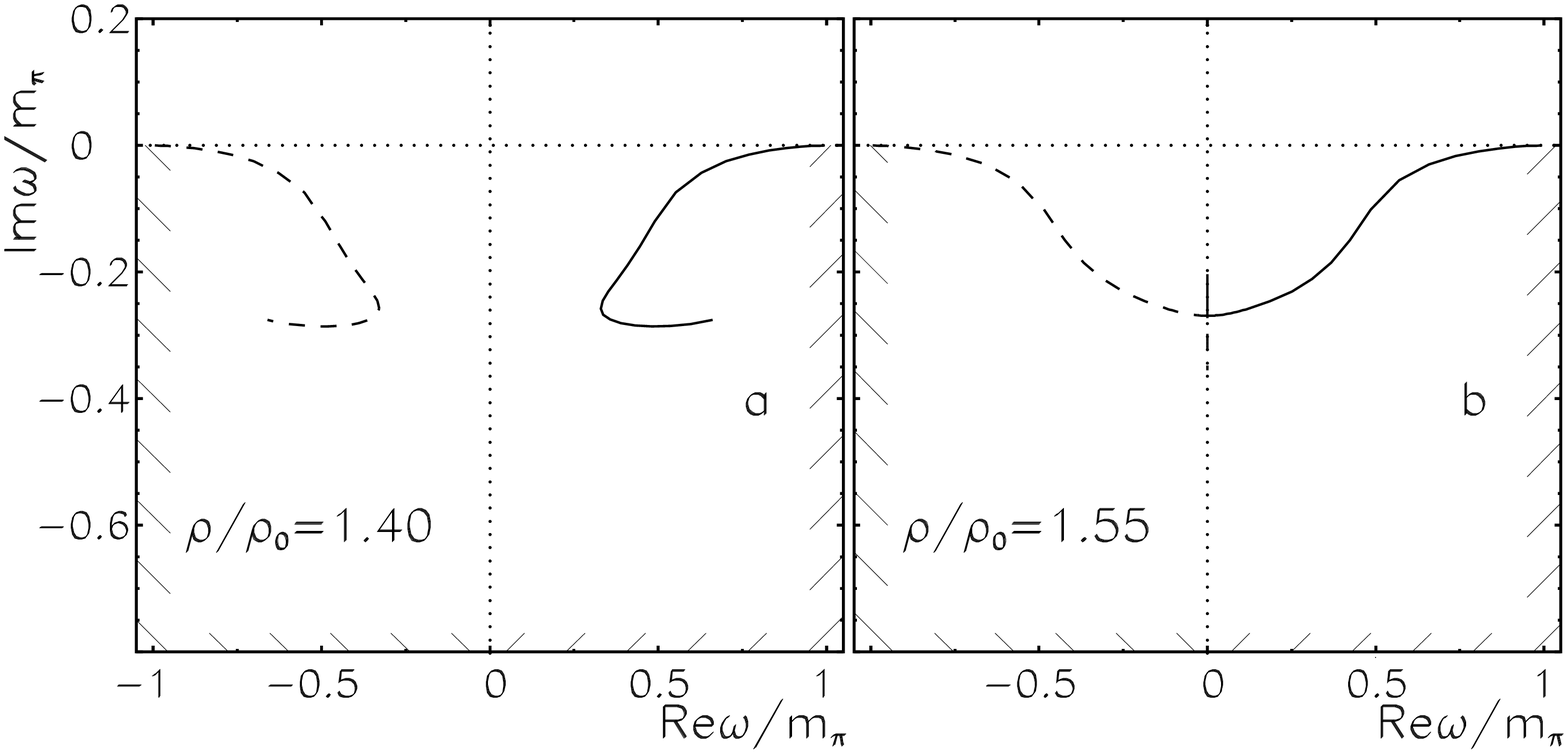,width=12cm}}
\centering{\epsfig{figure=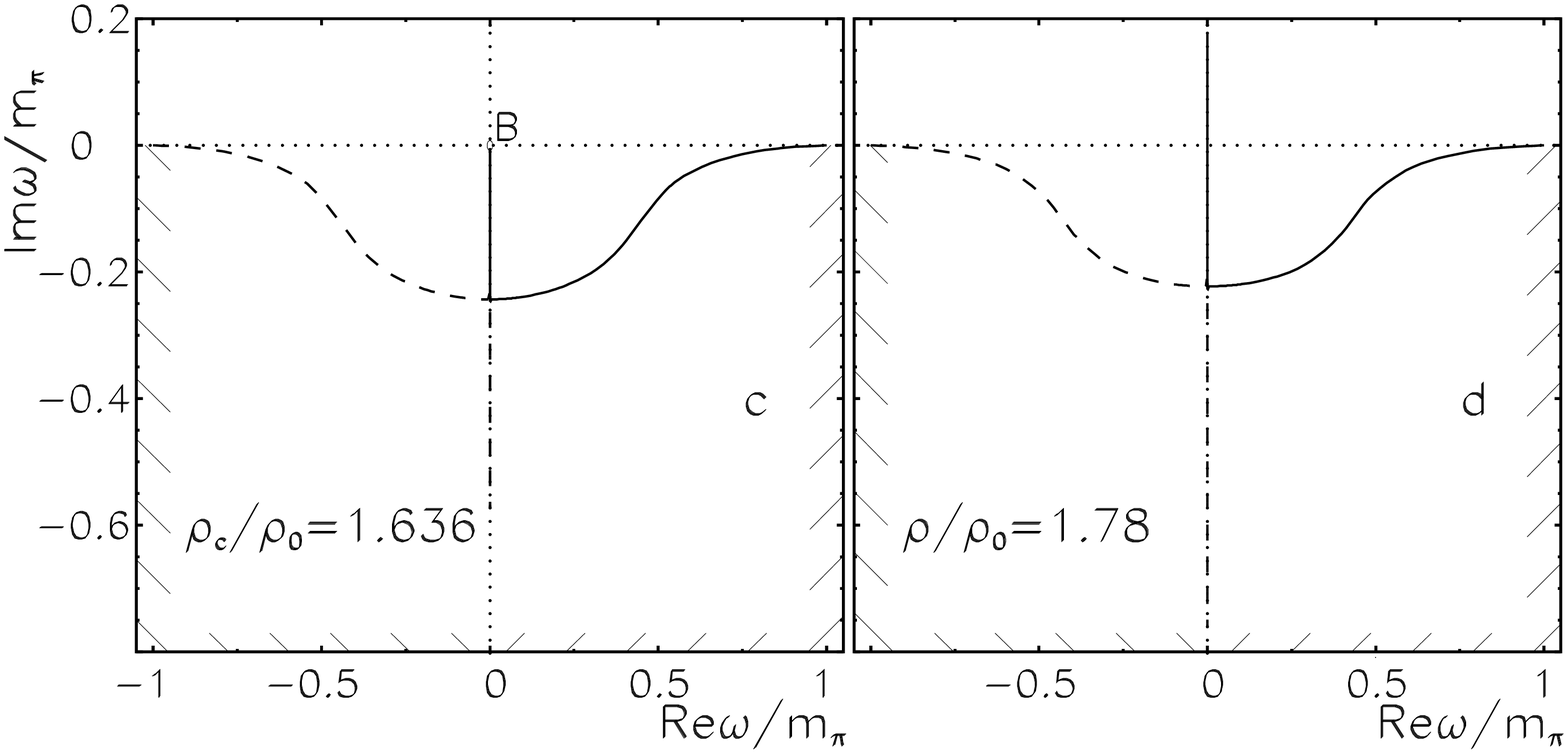,width=12cm}}
\caption{}
\end{figure}

%Fig.9
\begin{figure}
\centering{\epsfig{figure=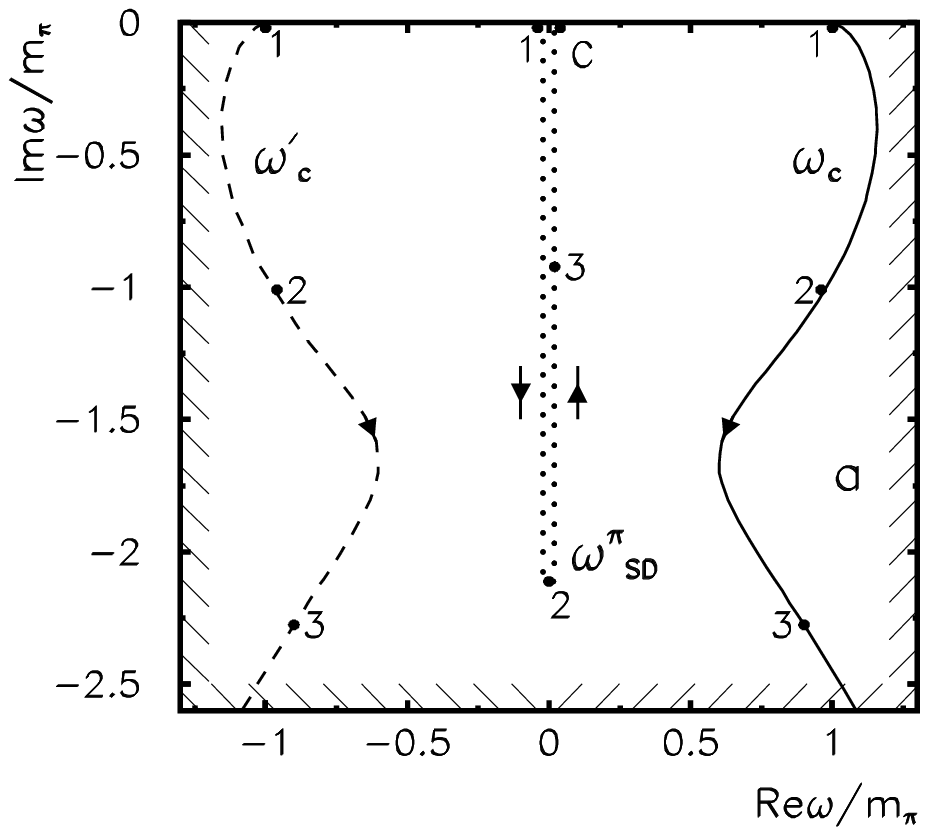,width=7cm}}
\centering{\epsfig{figure=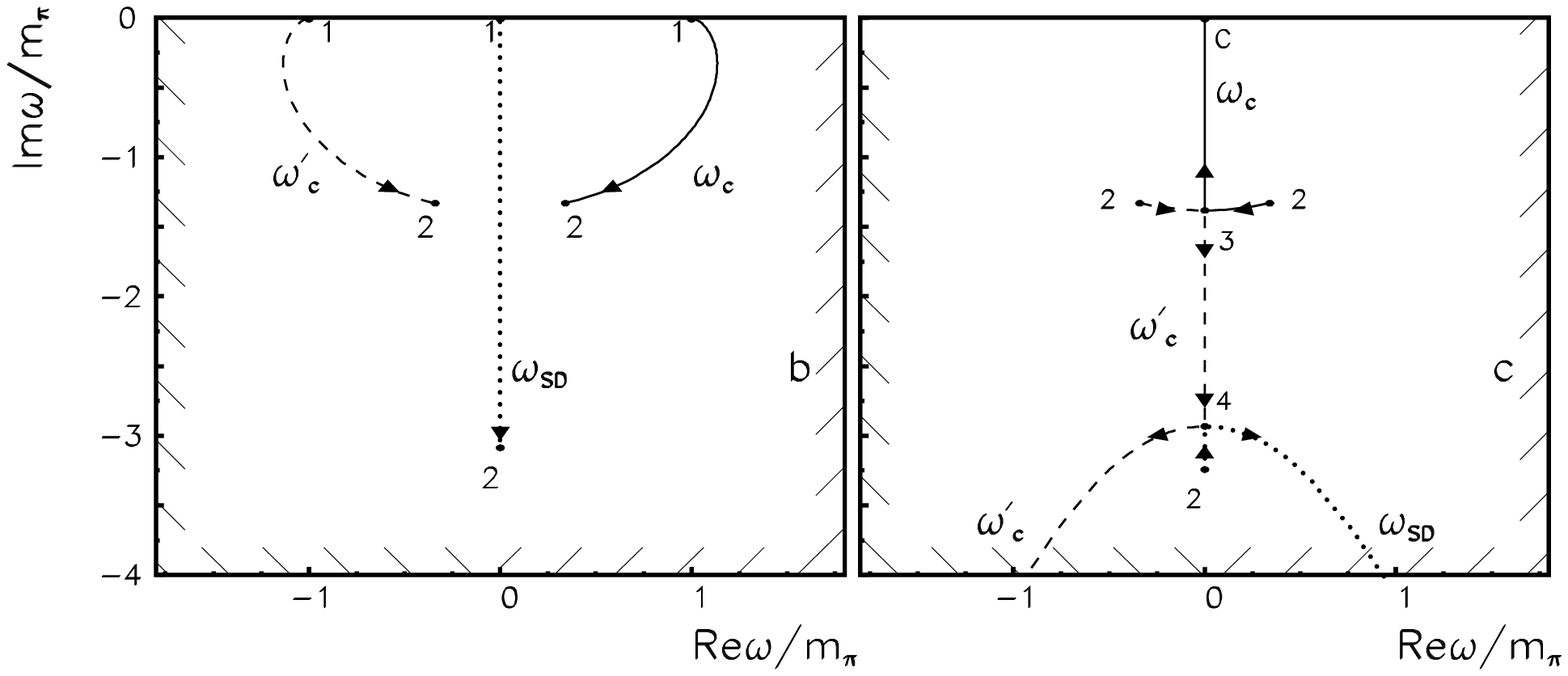,width=14cm}}
\caption{}
\end{figure}

%Fig.10
\begin{figure}
\centering{\epsfig{figure=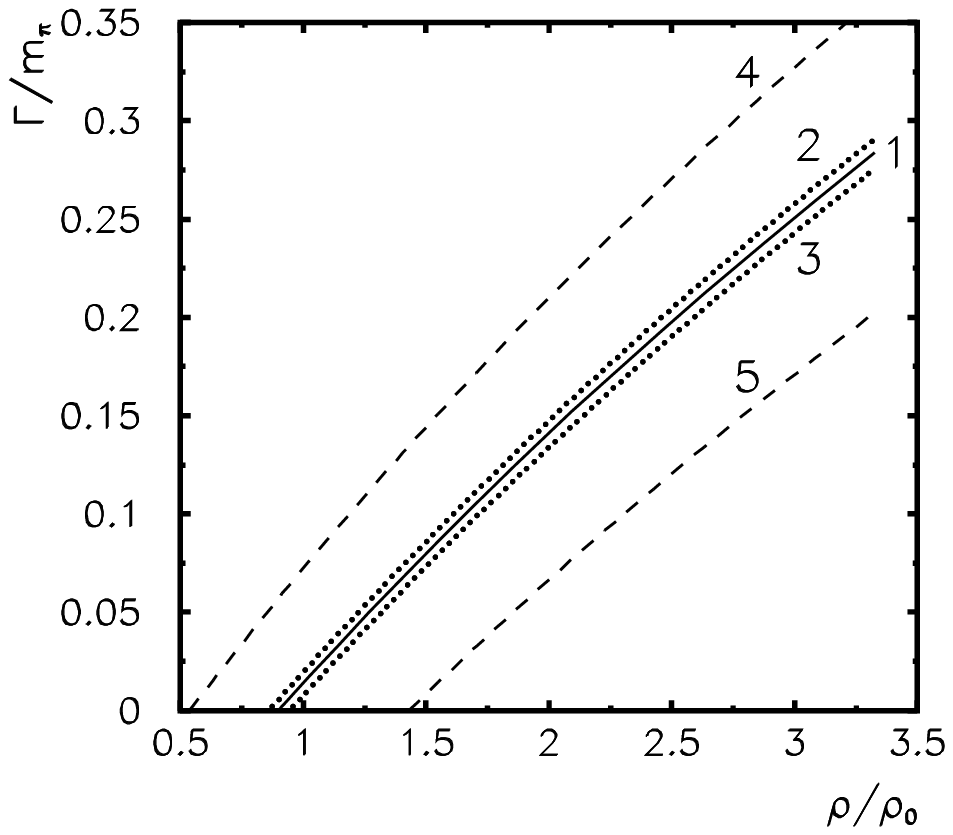,width=7cm}}
\caption{}
\end{figure}

\end{document}